\documentclass[acmsmall]{acmart}

\AtBeginDocument{%
  \providecommand\BibTeX{{%
    \normalfont B\kern-0.5em{\scshape i\kern-0.25em b}\kern-0.8em\TeX}}}

\setcopyright{acmcopyright}
\acmJournal{PACMHCI}
\acmYear{2021} \acmVolume{5} \acmNumber{CSCW2} \acmArticle{425} \acmMonth{10} \acmPrice{}\acmDOI{10.1145/3479569}

\received{January 2021}
\received[revised]{April 2021}
\received[accepted]{July 2021}

\usepackage{booktabs}
\usepackage[normalem]{ulem}

\newcommand{\system}{\textit{Deblinder}}
\newcommand{\embed}{Failure Report Embedding}
\newcommand{\hypo}{Hypothesis Panel}
\newcommand{\reports}{Failure Report Drawer}

\newcommand{\reps}{failure reports}
\newcommand{\rep}{failure report}

\newcommand{\para}[1]{\vskip 1mm\noindent\textbf{#1}~~}

\definecolor{blue}{HTML}{4285F4}
\definecolor{lightblue}{HTML}{e6effe}
\definecolor{red}{HTML}{EA4335}
\definecolor{yellow}{HTML}{FBBC04}

\newcommand{\n}[1]{#1}
\newcommand{\del}[1]{}

\definecolor{yellow}{HTML}{FBBC04}

\usepackage[most]{tcolorbox}
\tcbset{on line, 
        boxsep=3pt, left=0pt,right=0pt,top=0pt,bottom=0pt,
        colframe=white,colback=lightblue,  
        highlight math style={enhanced}
        }

\begin{document}

\title[Crowdsourced Failure Reports]{Discovering and Validating AI Errors With Crowdsourced Failure Reports}

\author{Ángel Alexander Cabrera}
\email{cabrera@cmu.edu}
\orcid{0000-0003-0348-3362}
\affiliation{%
  \institution{Carnegie Mellon University}
  \streetaddress{5000 Forbes Ave}
  \city{Pittsburgh}
  \state{Pennsylvania}
  \country{USA}
  \postcode{15213}
}

\author{Abraham J. Druck}
\email{adruck@andrew.cmu.edu}
\affiliation{%
  \institution{Carnegie Mellon University}
  \streetaddress{5000 Forbes Ave}
  \city{Pittsburgh}
  \state{Pennsylvania}
  \country{USA}
  \postcode{15213}
}

\author{Jason I. Hong}
\email{jasonh@cs.cmu.edu}
\affiliation{%
  \institution{Carnegie Mellon University}
  \streetaddress{5000 Forbes Ave}
  \city{Pittsburgh}
  \state{Pennsylvania}
  \country{USA}
  \postcode{15213}
}

\author{Adam Perer}
\email{adamperer@cmu.edu}
\affiliation{%
  \institution{Carnegie Mellon University}
  \streetaddress{5000 Forbes Ave}
  \city{Pittsburgh}
  \state{Pennsylvania}
  \country{USA}
  \postcode{15213}
}

\renewcommand{\shortauthors}{Ángel Alexander Cabrera et al.}

\begin{abstract}
AI systems can fail to learn important behaviors, leading to real-world issues like safety concerns and biases.
\del{Unfortunately, }\n{D}iscovering these systematic failures often requires significant developer attention, from hypothesizing potential edge cases to collecting evidence and validating patterns.
To scale and streamline this process, we introduce \n{crowdsourced} \textit{failure reports}, end-user descriptions of how or why a model failed, and show how developers can use them to detect AI errors.
We also design and implement \system{}, a visual analytics system for synthesizing \reps{} that developers can use to discover and validate systematic failures.
In semi-structured interviews and think-aloud studies with 10 AI practitioners, we explore the affordances of the \system{} system and the applicability of \reps{} in real-world settings.
Lastly, we show how collecting additional data from the groups identified by developers can improve model performance.
\end{abstract}



\maketitle

\section{Introduction}
AI systems deployed in the real world can have significant consequences when they fail, including safety issues like self-driving cars harming pedestrians~\cite{wakabayashi2018self}, and fairness concerns like inadvertently racist image labels~\cite{simonite2018comes}.
Developers need to be able to efficiently detect, validate, and fix systematic failures to improve the safety, equity, and overall performance of their systems~\cite{Rahwan2019}.

Unfortunately, discovering what systematic errors AI systems make presents developers with various challenges~\cite{Holstein2019, Kim2018, Amershi2019}. 
First, developers often have to sift through thousands of failure cases, many of which are random, one-off errors, to identify systematic failures.
This manual task can be prohibitively labor-intensive for individuals or small teams.
Second, once a developer discovers a pattern of failure, they have to validate their hypothesis by finding more evidence of the same behavior.
In common domains like images, there \del{is no single tool}\n{are few tools} for finding similar instances, making it hard to find more evidence.
Lastly, data scientists have to repeat this process to track and mitigate the various systematic failures a model may have.

To tackle these challenges, we define and formalize \n{crowdsourced} \textit{\reps{}}, text descriptions from end-users detailing how or why they think an AI system failed.
\n{Failure reports can, for example, describe a face recognition model that didn't detect a person who was outside, or a smartwatch that miscounted steps when a user was running on a treadmill.}
When collected at scale in a \textit{crowd auditing} process, \reps{} can be used to discover AI errors developers were unaware of and provide \del{additional} evidence to validate their hypotheses. 
Failure reports are inspired by the well-established concept of bug reports in software engineering~\cite{Bettenburg2008what}, but are different in a few significant ways. 
Primarily, since AI systems are often stochastic, black-box models, \reps{} \del{are unstructured and} have to be aggregated and further validated to uncover AI failures.
In this work, we discuss the similarities and differences between ~\reps~ and bug reports and the design of methods for collecting useful reports.
We also tested different collection methods and techniques and implemented an example report collection system using Amazon's Mechanical Turk platform.

\begin{figure*}
    \centering
    \includegraphics[width=\textwidth]{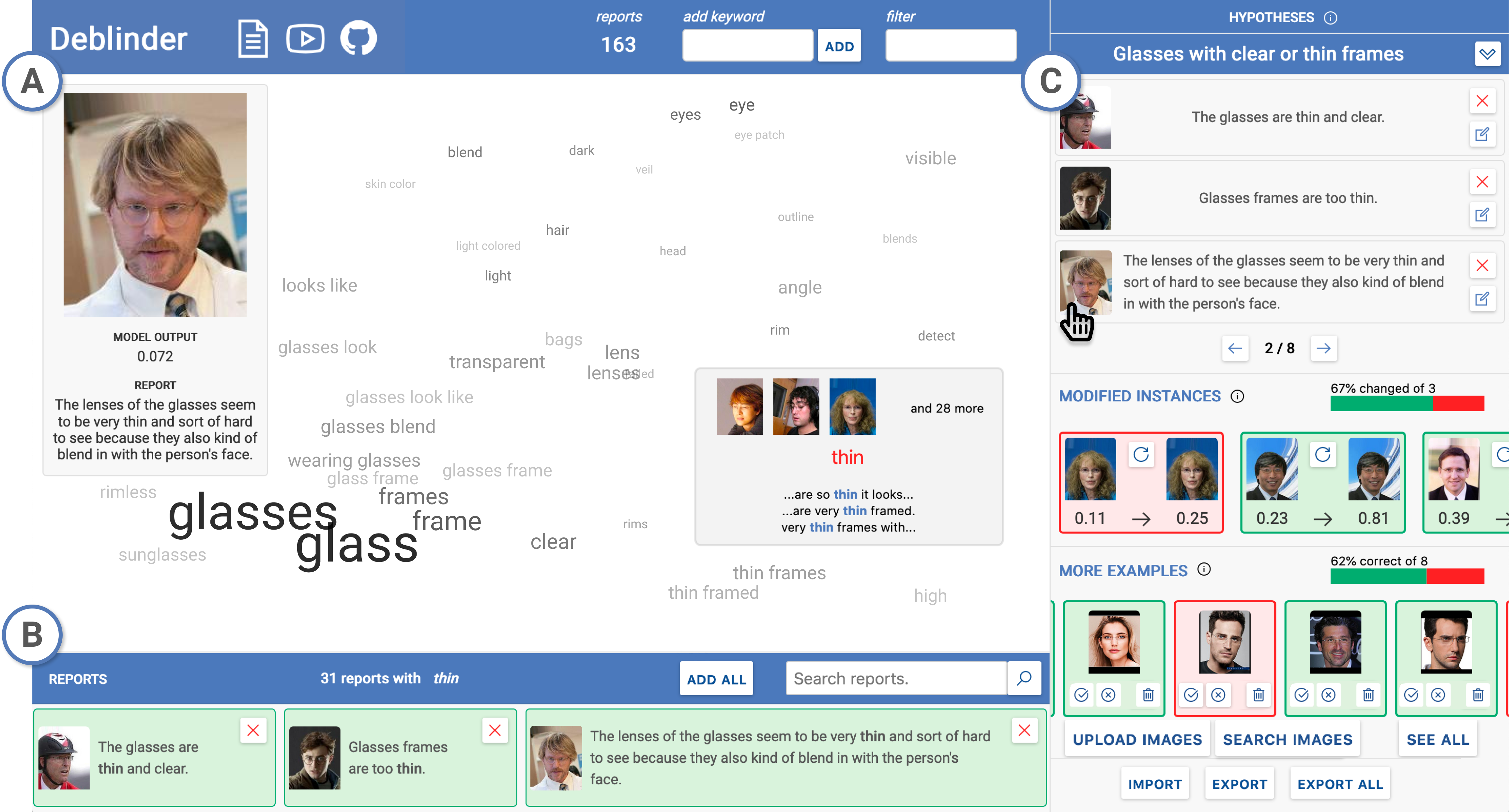}
    \caption{
    The~\system~system looking at~\reps~for a classification model that detects if a person is wearing eyeglasses.
    The descriptions of why each instance was misclassified are textual reports generated by crowdworkers.
    In the~\embed~\textbf{(A)}, a developer can explore high-level concepts extracted from the~\reps. 
    They can then search for or look through specific reports using the~\reports~\textbf{(B)}. 
    Finally, the developer can create hypotheses for blind spots in the~\hypo~\textbf{(C)} and test them by modifying instances or collecting additional data.}
    \label{fig:teaser}
\end{figure*}

Failure reports can describe AI failures, but collecting hundreds or thousands of free-text sentences leads us back to one of data scientists' original challenges: they still have to sift through countless \reps{} to find and validate patterns.
To address this issue, we designed and implemented a visual analytics system,~\system, that lets users explore hundreds or thousands of ~\reps~ to discover and test systematic failures.
The system's main interface is a visualization that aggregates~\reps~ to let developers find patterns of error.
Developers can then create hypotheses for failures they find in the report visualization.
\system~also provides two complementary features for validating failure hypotheses and assessing if they generalize - similar instance search and instance manipulation.
While~\system~is focused on image models, it is designed to be adapted to other domains like text and video data.

We explored the applications of failure reports and the~\system~system through semi-structured interviews and think-aloud studies with 10 AI practitioners.
We found that the process of creating and testing hypotheses for systematic failures mirrors developers' debugging process and that they found consistent failures with supporting evidence when using~\system.
The study also scoped the best uses of failure reports, including their limitation to domains where end-users can understand the input data and see the model output directly.
Lastly, we experimentally showed that a model retrained with failures discovered by study participants can improve model performance.

Failure reports add a useful strategy to developers' AI debugging toolbox. 
They complement existing algorithmic, visual, and crowd debugging systems by detecting and describing complex failures in deployment, like those developers may not have considered due to their own blind spots and biases~\cite{Holstein2019, Barocas2016, Selbst2019, Ramakrishnan2019}.
Visualizing crowdsourced \rep{}s continues the emerging theme of distributed or crowd sensemaking~\cite{Fisher2012, Goyal2015, Foong2017a, Kittur2014, Kittur2013}, which has been used to improve clustering~\cite{Chang2016, Andre2014}, summarize bug reports~\cite{Jiang2018a}, and learn model features~\cite{Cheng2015}.

In summary, our contributions are the following:

\begin{itemize}
    \item \textbf{\textit{Failure reports}, end-user descriptions of how or why an AI system failed.}
    We formalize a crowd auditing process for discovering AI failures based on \textit{\reps{}}, text descriptions of model errors from end-users.
    We explore the parallels between~\reps~and software bug reports, design methods for effectively collecting them, and implement an example collection method using Amazon Mechanical Turk.

    \item \textbf{\system, a visual analytics system for making sense of~\reps.}
    We designed and implemented a visual analytics system for synthesizing \reps{} developers can use to discover and validate AI failures.
    The system uses an interactive word-embedding visualization to aggregate and spatially organize the text reports.
    Developers can then create, track, and test hypotheses using \system{},  which provides two validation methods.

    \item \textbf{Evaluation of \reps{} and the \system{} system.} 
    In semi-structured interviews and think-aloud studies with 10 AI practitioners, we explored the real-world applications and limitations of \reps{}. 
    We found that developers discovered consistent, evidence-backed failures using~\system, and showed experimentally that collecting data from the discovered failures improved model performance. 

\end{itemize}

\section{Background and Related Work}\label{sec:related}

\para{Bug reports in software engineering.} 
Bug reports are an essential stage of the software engineering process.
A bug report generally consists of a description of the issue, steps to reproduce the problem, and supporting information like a stack trace~\cite{Bettenburg2008what, Zhang2015}, which software engineers can use to find and fix the failing code.
While at a high level failure reports are similar to bug reports, debugging AI systems present various new challenges.
Primarily, while a single bug report can be used to identify and fix a bug, various examples of the same issue are required to discover a systematic failure in an AI system.
Additionally, there is no ground-truth evidence like a stack trace~\cite{Schroter2010DoBugs} from faulty code in black-box AI systems - the only supporting evidence comes from failure reports and instances themselves.
Therefore, developers need numerous failure reports for the same issue and additional evidence to detect and validate AI failures.

While failure reports differ significantly from bug reports, we build on existing software engineering research on improving bug reporting.
Bug summaries can be helpful for quickly triaging bugs and finding similar issues~\cite{Rastkar2014AutomaticReports, Jindal2020}. 
For example, one method for improving reports showed how using crowd-elicited attributes could improve bug summaries~\cite{Jiang2018a}.
Visualizations, like the topic modeling approach by \citet{Yeasmin2014}, have also been used to improve bug reporting by summarizing bug repositories.
Lastly, finding duplicate bugs is also an active area of interest for reducing bug reports and finding more evidence for an issue~\cite{Sun2011, Sun2010, bettenburg2008duplicate}.
While we do not summarize or remove duplicate failure reports, we extract keywords and combine similar reports to provide a high-level view of the issues end-users describe.

\para{Error analysis for AI systems.}
We use the term \textit{systematic failures} to describe a group of instances sharing semantic features for which an AI produces the wrong output significantly more often than for the overall dataset.
Since there is no standard term in the literature for \textit{systematic failures}, we use it interchangeably with terms from existing work like \textit{blind spots}~\cite{Ramakrishnan2019} or \textit{unknown unknowns}~\cite{Bansal2018, Lakkaraju2017}. 

There are various \textit{algorithmic techniques} for discovering and characterizing systematic failures, ranging from fully automated to human-driven, crowdsourced techniques. 
Fully algorithmic strategies aim to automatically slice data to discover areas of error, for example
Lakkaraju et al.'s exploration-exploitation technique for clustering and discovering model blind spots~\cite{Lakkaraju2017} and Slice Finder, which splits data according to tabular features and uses the model's loss to rank their severity~\cite{Chung2019}.
These techniques require data that can be easily clustered or sliced, both of which are rarely available for data like images and videos.
By using human-generated failure reports, developers can find more nuanced and complex failures that are not defined by pre-existing features.

In addition to algorithmic methods, many \textit{visualization tools} exist for helping data scientists develop and debug AI systems.
These tools provide support across the entire ML development pipeline, from model tuning to error characterization.
For example,  Squares \cite{Ren2017} and MLCube \cite{Kahng2016} are visualizations for tracking models' performance across different dimensions, including class confusion and various performance metrics.
Visualization tools can also help developers debug models for specific types of error, for example, 
the What-If Tool \cite{Wachter2017}, FairSight \cite{Ahn2019}, and FairVis \cite{Cabrera2019} are visual analytics systems for auditing the fairness of AI systems.
These techniques, like the algorithmic methods, are limited to errors described using input features or model outputs.
By using text reports, our method can describe any human-defined failure. 

There are also visual systems specifically for exploring and characterizing model \textit{errors}.
Errudite is one such system specific for natural language processing (NLP) models~\cite{Wu2019}.
It uses a regex-based querying language for searching and replacing parts of text to discover and correct error hypotheses.
This technique works well for text data but requires users to already have hypotheses and does not generalize to domains like images.
AnchorViz is a polar-coordinates visualization that lets users create semantic anchors to visualize data instances across different concepts~\cite{Chen2018}.
This method requires users to manually label sentences with `anchors' or concepts, which does not scale easily.
Failure reports can surface initial hypotheses and supporting evidence for systematic failures.

Lastly, while not specifically designed for error analysis, data programming can be a powerful tool for discovering slices, or subgroups, of a dataset~\cite{Ratner2016, Heo2020}.
Data programming is a method of combining noisy labeling functions to train a classifier.
Developers can create labeling functions to quickly slice their data and discover systematic failures, similar to how MLCube~\cite{Kahng2016} and Slice Finder~\cite{Chung2019} help developers do subgroup analysis. 
Data programming can be a helpful validation tool for discovering more evidence for errors detected using \reps{}.

\para{Crowd auditing.}
Using crowdsourced human input has shown promise for discovering and characterizing AI models' systematic failures.
The first study to show that humans could effectively find AI failures was Beat the Machine, a fully human-driven technique that asked users to find examples for which an AI system failed, specifically for hate speech detection~\cite{attenberg2011beat}.
They found that humans could quickly find websites that the AI misclassified with very high probability. 
However, the authors did not take on the subsequent problem of aggregating the individually reported errors to describe and validate a model's systematic failures.

Crowds can also be used to establish the boundaries of AI behavior and prevent models from making harmful decisions.
\citet{Mandel2020} explored how to use the crowd to define acceptable AI behavior, created a rule-based interface to generate AI system constraints, and showed its efficacy in a real-world education domain.
\system{} and \reps{} can help identify the edge cases that would inform this type of crowd deliberation.

Recent work has explored how crowd input can be combined with algorithmic techniques to characterize AI failures.
\citet{Nushi2018} developed Pandora, a system that uses human and machine described clusters of data to derive a decision tree visualization of model errors.
Pandora is an effective method for finding features that correlate with or predict failure, visualized with a usable decision tree.
Like the clustering method by ~\citet{Lakkaraju2017}, Pandora is dependent on the clustering algorithm to find meaningful groups of failures. 
By using free text input, \reps{} 
can describe nuanced failures that may not be found with clustering.

\citet{Liu2020} introduced another method for detecting systematic failures, Patterned Beat the Machine (P-BTM), that extends BTM by using crowd workers to find more examples of unknown unknowns.
They ask crowd workers to provide initial labels for unknown unknowns, which are then used to train an \textit{expansion classifier} to find more examples in an unlabeled dataset.
Like clustering methods, P-BTM requires a classification model for any semantic concept, which they note is often challenging.
It also requires labeling initial unknown unknowns using a mechanism like BTM.
Crowd auditing with \reps{} uses the human labels to discover the initial failure instances \textit{and}, with a system like \system{}, surface similar instances that may be difficult for the expansion classifier \n{to find}.

Pandora and P-BTM are effective methods for detecting systematic failures that complement \reps{} and \system{ well.
Failure reports are better suited for finding semantic failures not present in a training or test set.
Since failure reports are text sentences, they can describe failures that involve actions or arbitrary semantic features, including those that are challenging to detect with a clustering or classification algorithm.
When collected for deployed systems, \reps{} can also detect real-world failures that are not present in a test set and labeled as an unknown unknown.
Pandora and P-BTM, on the other hand, can be more efficient and require less manual analysis since they partially automate the discovery and validation of systematic failures.
Developers can collect \reps{} and find initial systematic failures with \system{} that can be further validated using these hybrid, algorithmic methods.}

\section{Crowd Auditing With Failure Reports}

\subsection{Motivating Scenario}\label{sec:example}
To motivate the use of failure reports for detecting systematic failures and describe the process, we walk through an example scenario.

Kristen, an ML developer, has created an eyeglass detection system for use in airport customs.
For security reasons, people should not wear glasses when taking their picture at passport control.
Given a picture of someone's face, the system provides a binary output of whether or not the person is wearing glasses.
While the system has a high test set accuracy, Kristen wants to know if her model performs well in practice. 

At each photo kiosk, Kristen adds a text prompt that lets people write a report if the model fails to detect if they're wearing glasses or not.
After a month, she has collected thousands of~\reps, and uses~\system~to explore the reports and discover any systematic issues.
With this information, she collects additional data representative of those blind spots and retrains her model. 
The retrained model performs better on the test set, and she deploys the updated system.
She continues collecting reports to see if the blind spots persist or new issues arise over time.  

We collect~\reps~for the domain in this example, eyeglass detection, and use the results in our user study in Section~\ref{sec:user}.
Specifically, we trained a convolutional neural network (CNN) using PyTorch to classify headshots of people, with an overall accuracy of over 99\%.
We trained the model using images from the CelebA dataset, which has over 160,000 headshots of celebrities with labels for eyeglasses~\cite{liu2015faceattributes}.
Additionally, we collected reports and applied~\system~to the domain of image captioning (see Section~\ref{sec:captioning}) to show how the process and system generalize to other domains.

\subsection{Process Overview}
\begin{figure*}
    \centering
    \includegraphics[width=\textwidth]{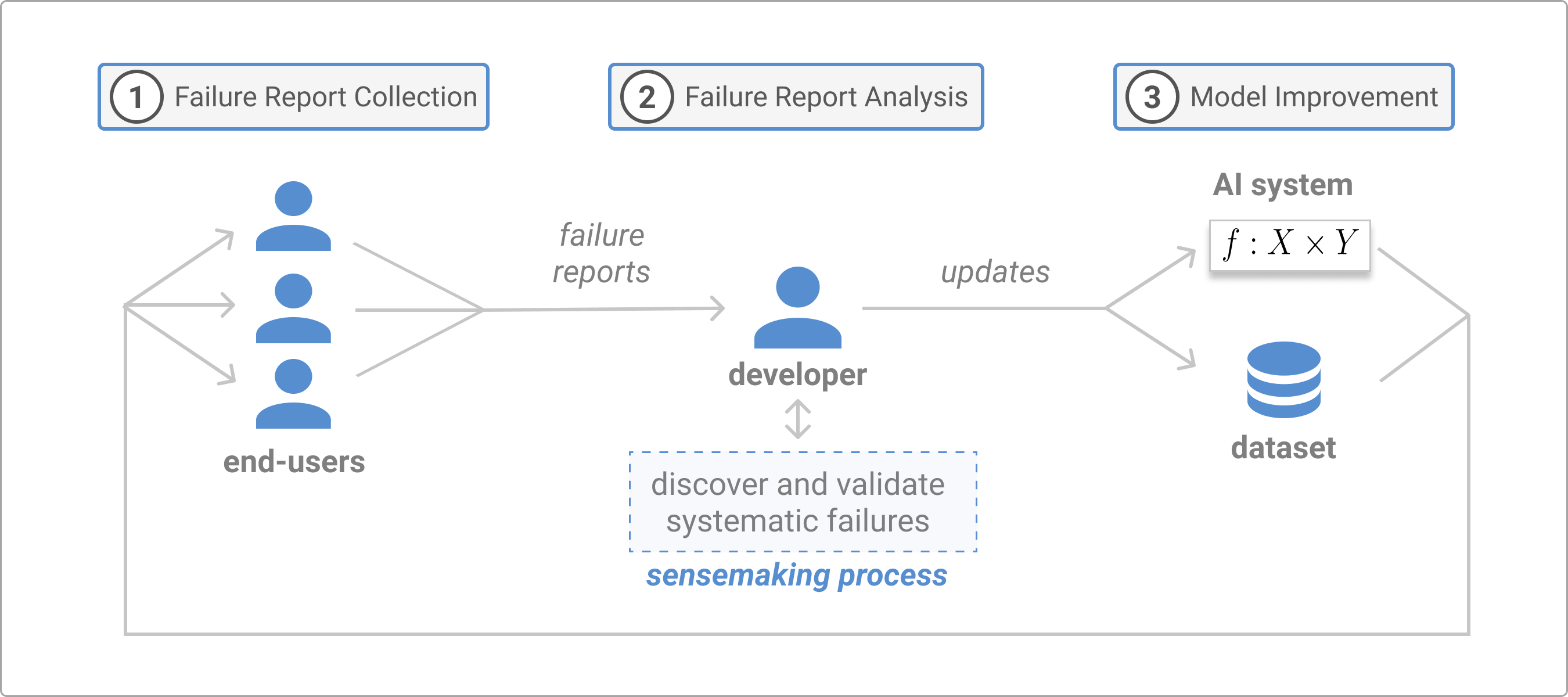}
    \caption{
    The iterative process of using failure reports to discover and mitigate AI models' systematic failures.
    \textbf{(1)} In the \textbf{Failure Report Collection} stage, end users observe and report model failures back to the developer.
    \textbf{(2)} Next, in the \textbf{Failure Report Analysis} stage, developers analyze the \reps{} to discover, describe, and validate hypotheses of potential failures.
    In this work, \system{} is used as the primary interface for this sensemaking process.
    \textbf{(3)} In the final \textbf{Model Improvement} stage, developers can use the discovered failures to collect additional data, tweak their models, and analyze the real-world impact of their systems.}
  
    \label{fig:process}
\end{figure*}

The crowd auditing process with \reps{} consists of three major stages, as can be seen in Figure~\ref{fig:process}.
It is an iterative and ongoing process, as end-users report newly discovered failures and developers update their model.
The three main stages are the following: (1) Failure Report Collection, (2) Failure Report Analysis, and (3) Model Improvement.

In the first stage, \textbf{Failure Report Collection}, end-users of an AI system describe why or how a model failed for a given instance. 
This stage can either be done during development to proactively find potential blind spots or be conducted in deployment to discover real-world problems.
The second stage is \textbf{Failure Report Analysis}, where developers have to make sense of the numerous text~\reps. 
In the last stage, \textbf{Model Improvement}, developers can use their validated insights to mitigate the real-world effects of model errors and improve their systems.

Using \reps{} to discover model blind spots provides unique advantages over standard manual and algorithmic methods.

Primarily, aggregated~\reps~can be a more cost and time-efficient option for finding systematic failures by surfacing commonly represented errors, a task that often requires significant manual labor. 

By including a more diverse set of users, this crowd auditing method can also detect and describe failures that developers had not considered due to their own blind spots and biases~\cite{Holstein2019}.
Additionally, if the process is applied to real-world domains instead of using crowd workers, the process can identify failures that are not present in the test set or that arise due to issues like dataset shift~\cite{Moreno-Torres2012, Recht2019}.
In the rest of this paper, we explore the challenges for collecting and analyzing \reps{}, and describe the design choices and technical solutions we took to address them.

\section{Failure Report Collection}
We define \textit{~\reps} as text \n{submissions} from end-users of AI systems describing a model's failure for a certain instance.
While~\reps~are conceptually similar to bug reports for software systems, their content differs substantially.
ML systems do not provide any additional insight into why an error occurred, like stack traces and logs in software systems~\cite{Zimmermann2010}.
A ~\rep~ and data instance is the only information reported back to the developer, and thus a report's entire value comes from the text.
Given the open-ended structure of ~\reps, developers can make various design choices when collecting reports, including the question they pose to end-users and the response format.
Here, we describe these design options and the specific choices we made for this work.

\subsection{What is a Failure Report?}
Failure reports are free-text responses from end-users describing certain features of an instance.
While we tested other types of reports, specifically short `tags' describing an error, we found that descriptive sentences were necessary to capture nuance like actions and complex image features. 
Free-text sentences are also a low barrier of entry for end-users that do not require any additional training, instruction, or domain knowledge.
Beyond being a short sentence, there is considerable freedom in what developers can ask end-users to describe in their~\rep.
Through testing in different domains and results from our user study, we found that the type of report a developer should collect depends on a model's domain.

In complex tasks with ambiguous ground truth or performance measures, it is often most useful to collect responses of \textit{how} a model is failing.
For example, in a system that automatically generates captions for an image, collecting descriptions of how the captions are wrong (e.g., the dog is holding a frisbee although the caption says it is a tennis ball) can provide model developers with a mental map of the kinds of mistakes the model is making.

In more well-defined domains like classification, collecting responses of \textit{why} a model is failing is often more useful since the type of error is apparent (the wrong class).
For a system that detects if someone is wearing glasses, for example, it is often clear what the problem is (e.g., the glasses were not detected) but not why it happened (e.g., the glasses had thin frames, or the person was looking to the side).
Descriptions of what aspects of an instance end-users think caused the algorithm to fail can provide the context needed to help developers find systematic errors.

In this work, we used different techniques for the two image domains we analyzed.
In the eyeglass detection example, we collected \textit{why} descriptions since the model failures are clear, and for image captioning we collected \textit{how} descriptions to discover the ways in which captions were wrong.

\begin{figure}
  \centering
  \includegraphics[width=.5\textwidth]{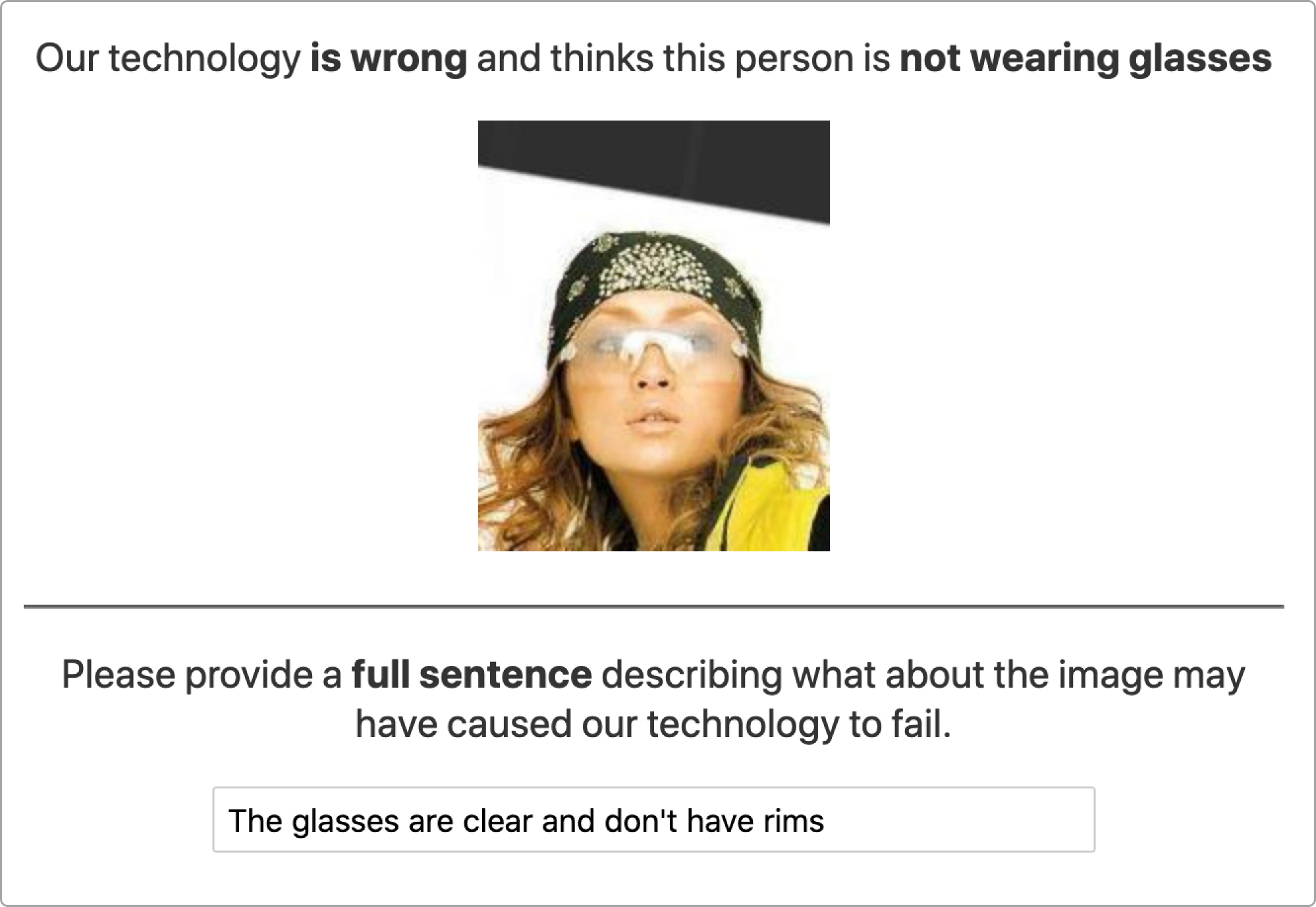}
  \caption{The web-based interface used to collect ~\reps. 
  We provided Amazon Mechanical Turk (AMT) \n{workers} with a series of 3 images that a model had failed for. 
  In this example, the AMT worker was given a false negative error from the eyeglass detection model which they attributed the error to the person's glasses being rimless and reflective.
  These reports were used in~\system~for our user study and evaluation.}
  \label{fig:mturk}
\end{figure}

\subsection{Methods for Collecting Reports}\label{sec:collecting}
In addition to \textit{what} developers ask end-users to describe, developers also have to decide \textit{how} they will collect reports.
We define two primary methods for collecting ~\reps: from end-users in a deployed setting or workers on a crowdsourcing platform.
For a deployed AI system,~\reps~can be collected in the same way software bug reports are - when users see that a model's output is wrong, they can submit free-response text reports through a reporting interface.
This can be implemented as a dedicated error reporting UI, or integrated into the existing support and feedback features of a product.
Crowdsourcing platforms are a viable alternative that can be used to gather ~\reps~ if end-users do not typically see a system's outputs or developers want immediate results.
The primary difference between the two methods is that crowdsourcing platforms require the developer to have a set of labeled failure instances to show crowdworkers, while in deployed systems end-users discover the model failures themselves.

To run controlled experiments and reproduce our results, we used the crowdsourcing strategy in this work and created a web-based interface for end-users to describe how or why a model failed for an instance. 
We used Amazon Mechanical Turk (AMT) to collect~\reps~using a web interface we created seen in Figure~\ref{fig:mturk}.
We showed each worker three instances, and for each instance asked the worker for a description of either how or why, depending on the domain, the model had failed.
The task took an average of 5 minutes, and we paid participants \$1, the equivalent of \$12 per hour.
We used the system to collect~\reps~for both the eyeglass detection and image captioning models, collecting 163 and 55~\reps~for each domain respectively.
This report collection interface is a proof of concept for this work - real-world systems would implement their own report collection platforms and processes specific to their product.

Since AMT is generally representative of US internet users (with some small, consistent variation~\cite{Ipeirotis}), it provides a good approximation of the type and quality of reports that would be gathered in public-facing AI applications.
To control AMT responses' quality, we limited participants to those who had an approval rating of over 98\% with over 500 completed tasks.
We found the responses to be of high quality generally, 
corroborated both by direct comments by study participants (AI developers) and the \n{consistency and impact} of the failures they discovered (Section~\ref{sec:user}).
One developer commented that ``there's no way this is Turker data'' due to the reports' high quality, and all developers found consistent, evidenced blind spots from the AMT~\reps.
Future work could explore methods for incentivizing good reports and study the differences between failures discovered by the general public and data scientists.

\section{Failure Report Analysis}
Failure reports provide the core insights into model errors, but to discover valid systematic failures developers have to extract aggregate patterns and supporting evidence from hundreds or thousands of text reports.
To support this process, we present the design and implementation of a visual analytics system,~\system, that aggregates \reps{} in a word embedding visualization and lets developers create and test their blind spot hypotheses. 

There is a rich literature of visual analytics approaches for exploring large corpora of text~\cite{Kucher2015}.
This text analysis is often described \n{as \textit{sensemaking}, the process by which} people organize large datasets to understand and validate patterns~\cite{Weick1995, Pirolli1999, Pirolli2005}
The sensemaking framework specific to data analysis by \citet{Pirolli2005} has been used to derive the design requirements for many visual analytics systems.
One such system is Jigsaw, a visualization system for exploring large text datasets~\cite{Gorg2013}. 
The authors used algorithmic methods to summarize, aggregate, and organize documents in a visual interface.
Other systems using the sensemaking framework include the Aruvi system for analytical reasoning ~\cite{Shrinivasan2008} and Apolo for exploring large networks~\cite{Chau2011}.

Like these existing visual systems, we derive design challenges for \system{} using the established sensemaking framework by~\citet{Pirolli2005}.
Sensemaking consists of two processes - information foraging, where an analyst finds evidence and forms initial ideas, and synthesis, where the analyst creates and tests formal hypotheses.
The three primary design challenges we define come directly from the central stages of the sensemaking process, the [R1] evidence file, [R2] schema, and [R3] hypotheses.
We tailor the challenges specifically for text analysis and AI failures.

\subsection{Design Requirements}

A visual analytics system should address the following challenges for \n{analyzing}~\reps~:

\begin{itemize}
    \item[\textbf{R1.}] \textbf{Extract, filter, and search for high-level concepts.}
    \newline
    Text-based ~\reps~ should be summarized in a way that scales to thousands of reports.
    Developers should be able to quickly see an overview of high-level concepts derived from raw reports.
    Developers should also be able to apply their own domain knowledge of possible failures by filtering and searching for specific concepts and reports.
    
    \item[\textbf{R2.}] \textbf{Meaningfully organize concepts and~\reps{}.} 
    \newline
    To develop and validate hypotheses, it is useful to know what concepts and reports are most similar to each other.
    This context allows users to brainstorm hypotheses that include a variety of similar concepts and find reports that may fit into their existing hypotheses.

    \item[\textbf{R3.}] \textbf{Create, manage, and validate multiple hypotheses.} 
    \newline
    Developers should be able to create, track, and test formal hypotheses of systematic errors.
    They should be able to create and name hypotheses with supporting evidence, ~\reps, and testing instances.
    When a developer has a hypothesis of a certain systematic failure, they should have methods for validating their hypothesis.

\end{itemize}

\subsection{System Design}
The~\system~interface, seen in Figure~\ref{fig:teaser}, is primarily composed of two views, the~\embed~(Figure \ref{fig:teaser}A) and the~\hypo~(Figure \ref{fig:teaser}C). 
The~\embed~is the first point of entry and provides an interactive, aggregated visualization of~\reps. 
Developers explore and select reports from this view to generate hypotheses in the~\hypo. 
In the~\hypo, they then create and collect evidence for hypothesized systematic failures.
Developers can then use two different strategies to validate their hypotheses: modify instances to see the model's new output, or collect additional instances to see if the systematic failure generalizes to similar instances.

\subsubsection{Making Sense of Failure Reports}

\begin{figure}
    \includegraphics[width=.75\textwidth]{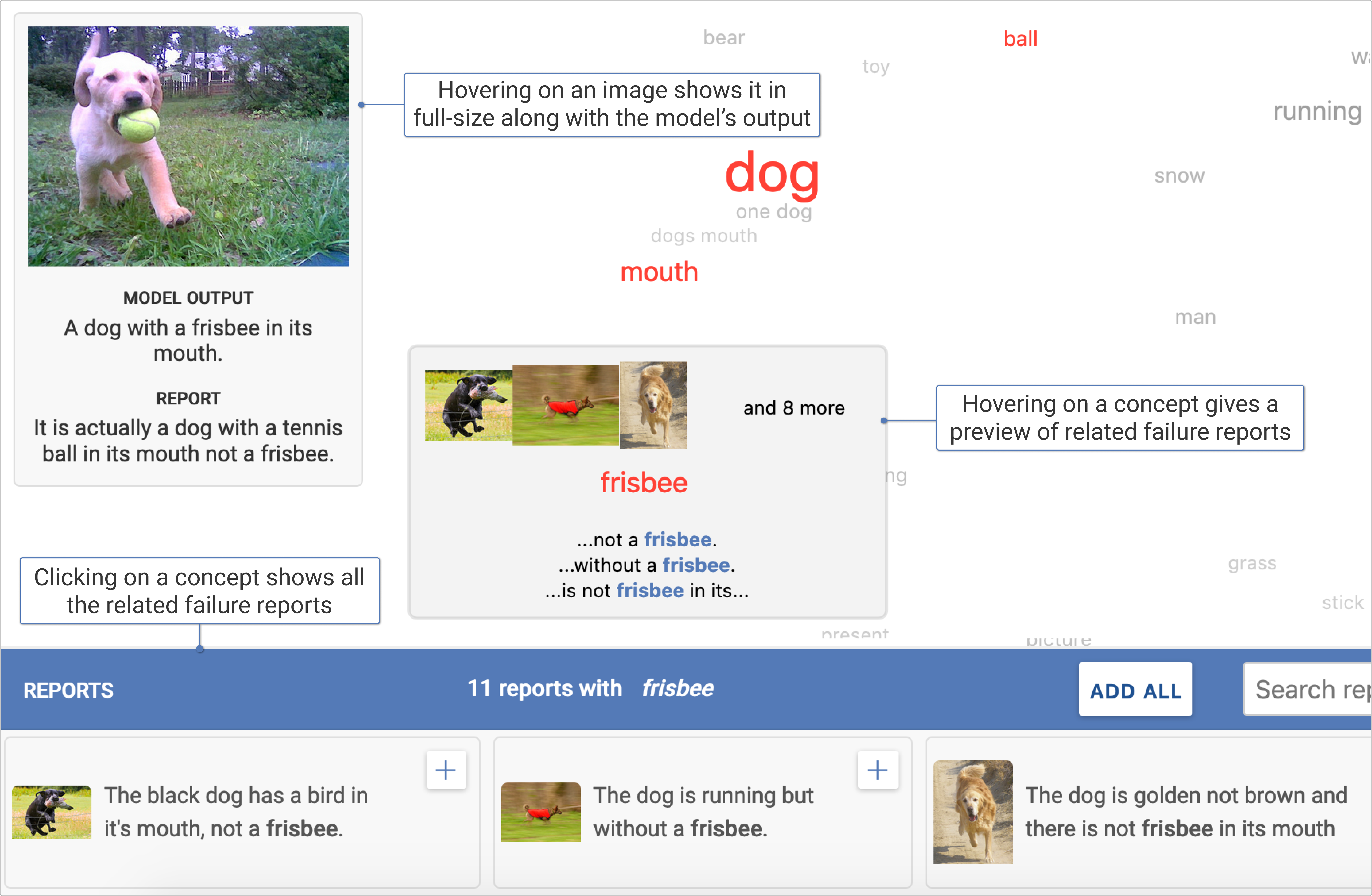}
    \caption{The \textbf{\embed} view used to explore and derive insights from reports, shown with reports for the image captioning model. 
    Concepts are extracted from reports and displayed using word embeddings.
    When a user hovers over a concept, it shows a few example reports and instances. 
    Clicking on a concept expands all the reports in the reports panel, allowing users to add them to their hypothesis. 
    Hovering over any instance in the interface shows the full instance in the upper left corner with the model's prediction.}
    \label{fig:embedding}
\end{figure}

The~\embed~is the primary interface with~\reps.
The collected reports are free-response sentences semantically describing either how or why an end-user believes the AI system failed for a given instance.
To give users a high-level view of which types of errors are happening most often, we extract \textit{concepts} from the reports \textbf{[R1]}, key phrases from the reports representing commonly mentioned terms. 
These concepts are aggregated, counted, and displayed in the~\embed~visualization, an example of which can be seen in Figure~\ref{fig:embedding} for the domain of eyeglass detection.

To extract concepts from the reports, we use RAKE (Rapid Automatic Keyword Extraction), a domain-agnostic keyphrase extraction algorithm~\cite{Rose2010}.
The extracted high-level concepts are then shown in a two-dimensional word embedding visualization.
To create the visual embedding, we use word vectors and dimensionality reduction.
For each extracted concept, we calculate its embedding vector using the GloVe (Global Vectors for Word Representation) algorithm~\cite{pennington2014glove}, which gives us a quantitative measure of how similar each concept is to each other.
To project the 300-dimensional GloVe vectors of each concept into a two-dimensional plot, we use the UMAP dimensionality reduction algorithm, which maintains the locality of points in the reduced space~\cite{mcinnes2018umap-software}.  

We decided to use a visual embedding instead of a table or word cloud to fulfill the proposed design requirements. 
A central requirement for the~\reps~visualization is showing the semantic similarity between reports \textbf{[R2]}.
This is helpful because it can help users find similar concepts that may not share the same text. For example, in the eyeglass detection model, the concepts \textit{frames} and \textit{rims} are closely related and often interchangeable.
A visual embedding groups similar concepts like these in space - in the same eyeglass domain, the concepts for \textit{glasses}, \textit{frames}, and \textit{lenses} are located close to each other. In contrast, the concepts for \textit{helmet} and \textit{hat} are tightly coupled in another area of the embedding.
This locality allows users to quickly see the relationship between concepts and explore reports that are similar in meaning.
The visual embedding medium is also interactive and allows developers to quickly go from concepts to individual reports, which would not be easily accomplished with a table.

To improve the usability of the~\embed~we use empirical findings from visualization research.
For each extracted concept, we calculate how many times it is mentioned across all ~\reps, and set the size and opacity of its text accordingly.
This both gives developers an idea of what the most common reports are \textbf{[R2]} and helps deal with the scalability issue of too many reports crowding the embedding and reducing legibility. 
Additionally, research has found that keyword discovery, a task similar to exploring concepts, is best aided by changes to the spatial layout and font of the text~\cite{Felix2018}.
The final usability feature we include is a form of semantic zooming~\cite{Bederson2000}.
As users zoom into sections of the~\embed, we rescale the font size and opacity, spreading out the phrases and making smaller concepts easier to see.

After exploring high-level concepts, users next have to look at specific reports and instances to support their hypotheses.
When a user hovers over a concept, a preview box appears around the phrase.
It includes a few example instances, the number of reports with that concept, and excerpts from a few reports giving context to how end-users are using the concept.
When the user clicks on the concept, it pins the phrase and expands all the related reports in the \textit{reports panel} (Figure~\ref{fig:teaser}B).
The reports panel shows all the instances and full ~\reps~ for the selected concept and allows users to add the reports to their hypotheses.

Lastly, it is often the case that a developer has domain knowledge about the model they might want to apply.
Towards this end, we include several utility features in~\system~\textbf{[R1]}. 
We let users add new keywords to the~\embed, which are placed with the appropriate scale and coloring where they semantically belong.
There is also a filtering feature that dynamically highlights in the embedding any concept the developer writes.
Users may also want to find key phrases or words in the ~\reps~ themselves, which we enable with the search functionality in the reports panel.
With these utility features, developers can correlate their intuitions and findings with the ~\reps~ they gather. 

\begin{figure*}
    \centering
    \includegraphics[width=.75\linewidth]{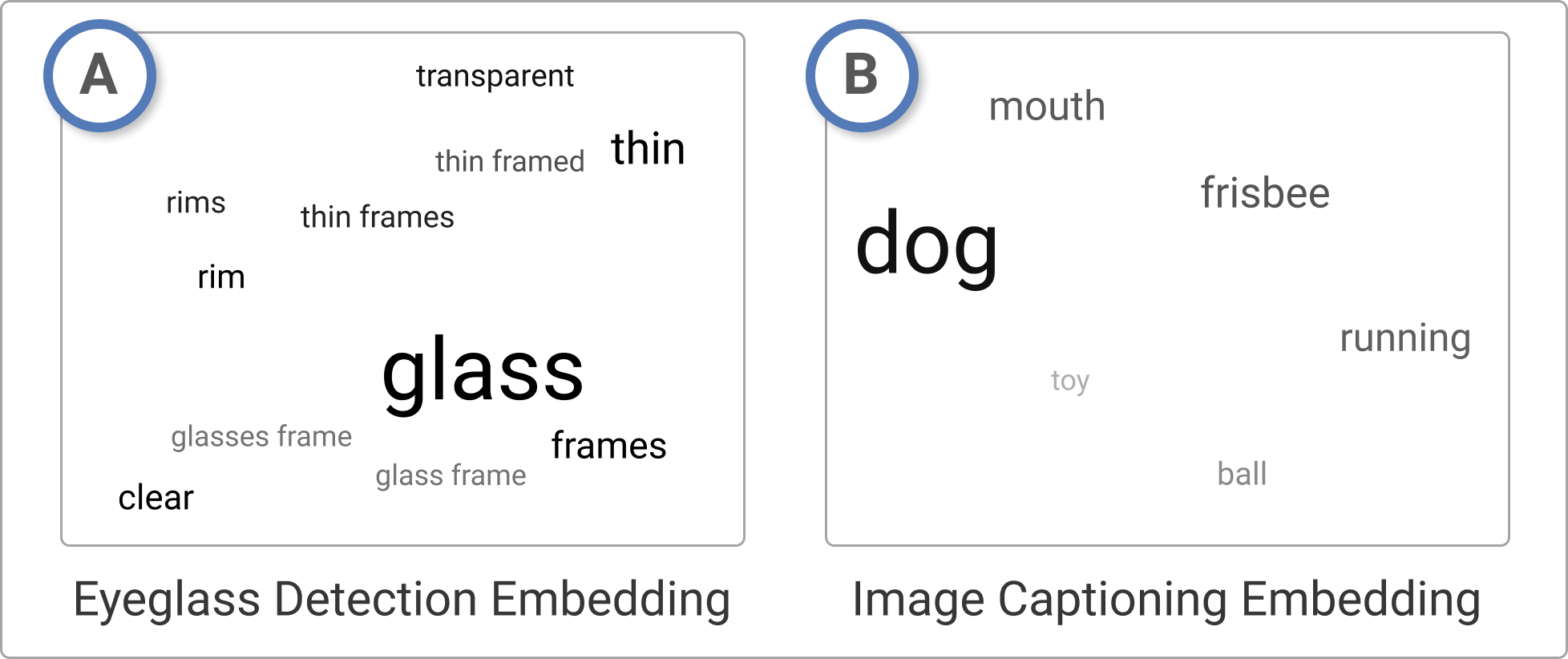}
    \caption{Example areas of the~\embed~for the eyeglass detection and image captioning domains.
    In the eyeglass detection embedding, reports for \textit{thin}, \textit{transparent}, and \textit{rims} are close to glasses, priming the hypothesis for clear or thin glasses frames.
    For image captioning we see the concepts for \textit{frisbee} and \textit{running}, hinting at the common error of captions mentioning a nonexistent frisbee for standing dogs.
    }
    \label{fig:use-cases}
\end{figure*}

\subsubsection{Creating and Tracking Hypotheses}

Once developers have generated initial hypotheses using the~\embed, they can use the~\hypo~ (see Figure \ref{fig:teaser}C) to create, track, and further test systematic failures.
Developers can name and track multiple hypotheses consisting of representative ~\reps, and use two validation methods to test their initial ideas.

\begin{figure}
    \centering
    \includegraphics[width=.5\textwidth]{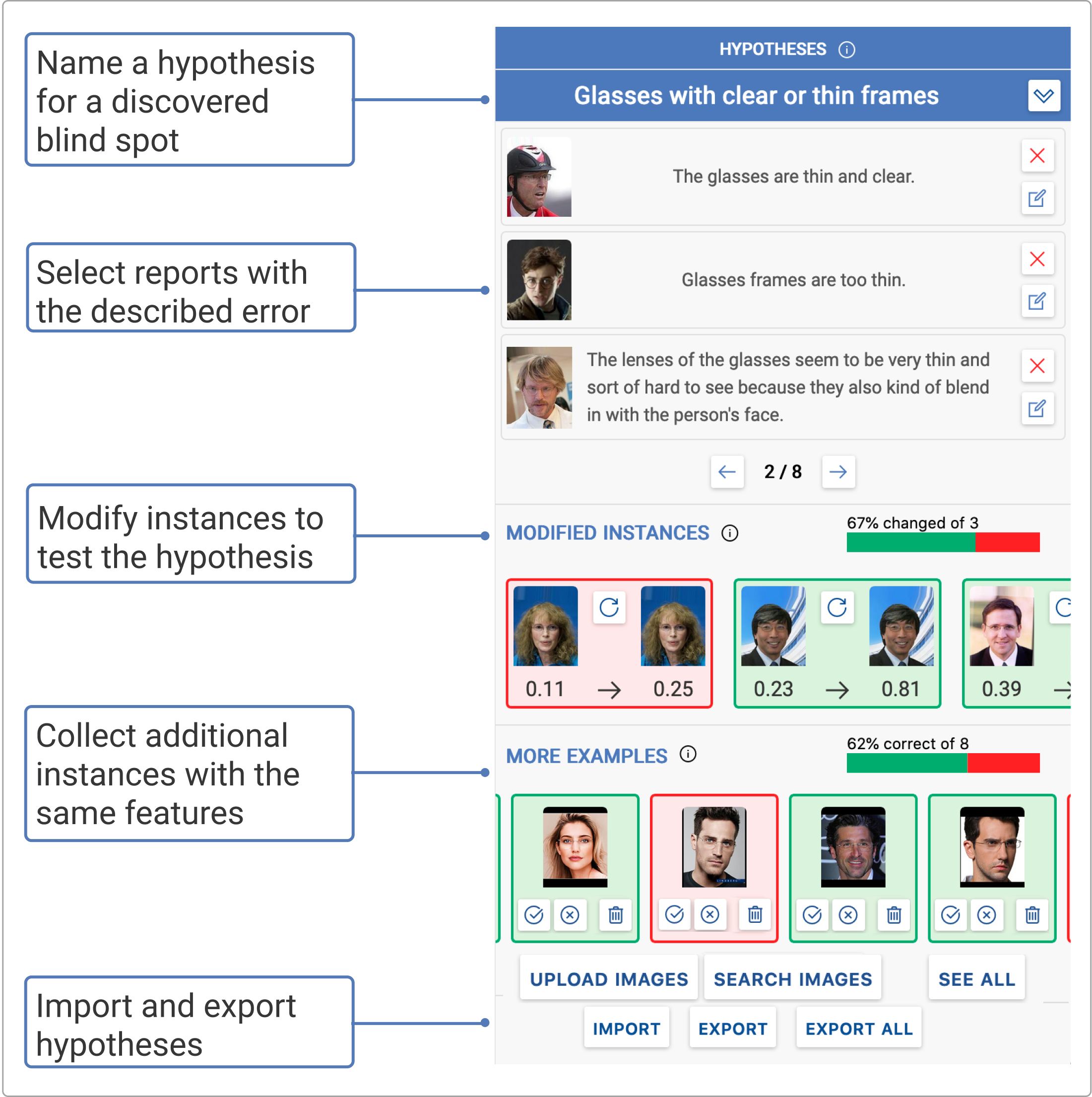}
    \caption{The~\textbf{\hypo}~ developers use to create, track, and test hypotheses.
    Developers can name hypotheses according to the features describing a systematic failure.
    For each hypothesis, developers add reports that match the hypothesis reason.
    They can then test the hypotheses with the modified instance and additional example strategies.
    Lastly, users can import and export hypotheses files for collaboration and integration into existing ML development tools.}
    \label{fig:hypotheses}
\end{figure}

The first step developers take is creating and adding evidence to a hypothesis, which is done in the~\hypo.
At the top of the~\hypo~developers can name the current hypothesis according to a specific type of error.
Using the dropdown menu, they can switch between different hypotheses or create a new hypothesis \textbf{[R3]}.
An example of the~\hypo~can be seen in Figure~\ref{fig:hypotheses} for the domain of eyeglass detection.

When a hypothesis is selected, the first section of the panel shows the ~\reps~ the developer has added to the hypothesis.
These reports are instances end-users reported due to the failure reason described by the hypothesis.
When developers hover on each ~\rep~, the concept's related reports are highlighted in the~\embed, allowing the user to find potential new concepts to explore and expand their hypotheses \textbf{[R3]}.

\subsubsection{Validating Hypotheses}
While~\reps~provide strong evidence of a model's systematic failures, they are not sufficient evidence to conclude that the issue generalizes.
Failure reports are an incomplete source of evidence for two primary reasons.
The first is that ~\reps~ only show us instances for which the model is wrong.
In practice, it may be the case that the model is correct for most instances described by a hypothesis.
In the eyeglass detection domain, it may be that most people with clear glasses are correctly classified at a similar rate as people with dark-rimmed glasses, despite various~\reps~mentioning clear glasses.
The second way in which ~\reps~ are insufficient evidence is in testing how plausible end-user reasons for model failure are. 
The reasons for failure reported by end-users are subjective and may not be real causes of a model's failure. 
Using the same example as above, it could be the case that the images of people with clear glasses were incorrectly classified because they were all worn low on the face, confounding the provided reason of clear glasses.

To address these issues, we introduce two methods in~\system~that allow users to test their hypotheses and validate how viable each one is: additional instance search and image manipulation.
The first testing method available in~\system~is running the model on additional instances described by a hypothesis.
By looking at the model's performance on additional instances, developers get a better approximation of how their model performs on a given blind spot compared to the overall dataset.
While the most straightforward way to find additional examples would be to search through the training and testing sets, this method presents a couple of significant difficulties.
First, finding images that match an arbitrary human-defined feature, for example, `a person with clear glasses looking sideways', is a challenging and generally unsolved problem.
Second, if the types of model failures reported by end-users are not present in the original dataset, additional instances of the reported issues would not be found with this strategy.

Due to these difficulties, we include a different method for discovering new instances inspired by Beat the Machine~\cite{attenberg2011beat}: we allow developers to find additional data in the wild \textbf{[R3]}.
Search engines are often able to find data for arbitrary descriptions, so~\system~embeds a search feature for finding images using the Flickr API, which developers can use directly from the interface to find and upload images for a given search term.
\system~also allows developers to upload new instances developers find using other means.
The AI system is then run on the new instances to get the model's output, and developers can indicate whether the model is correct or incorrect.
A percentage bar tracks the model's accuracy on the new instances, giving developers an approximation for how well their hypotheses generalize. 

This technique can also be used for the last stage of the process, Model Improvement, by collecting additional training data.
As we show in Section~\ref{sec:expirement}, the additional images collected as evidence can be added to the training set to improve the model's performance.
To be model and framework agnostic, \system{} does not have a feature for directly retraining the AI system but lets users export the discovered images to be included in future training \n{sets}.

The second \n{validation} technique available in~\system~lets developers directly test their hypotheses by modifying an instance and looking at the model's changed output.
For a reported or additional instance, a developer can download the instance, modify it, and re-upload it \textbf{[R3]}. 
Developers can use any tool or system they are comfortable with to modify the instance, ranging from photo manipulation software to algorithmic manipulations.
This technique takes inspiration from \citet{Cabrera2018}'s web-based interface that allows users to remove features of an image and see the model's new output.
However, we decided not to include any predefined image manipulations to allow developers to use techniques they are familiar with to modify the image in any way.
For the example hypothesis above of clear glasses, the developer may download an image of someone with clear glasses and darken the glasses frames.
They can then upload the edited image and see the model's output.
The model output for both the original and modified instance are presented to the developer, who indicates whether the model's output changed as expected.
As with the additional image validation, a percentage bar tracks how many of the modified instances changed as expected, giving the user a heuristic for how valid their hypothesis is.

\subsection{Implementation}
\system~is a web-based system built using Svelte\footnote{\url{https://svelte.dev/}} and D3\footnote{\url{https://d3js.org/}} for the front-end, and Flask\footnote{\url{https://flask.palletsprojects.com/en/1.1.x/}} for the server, both of which can be run locally.
The Flask server hosts the machine learning model, which is used to get the model output for additional examples and modified images.
The resulting hypotheses and evidence can be saved, exported, and shared.
\system~will be open source and can be adapted to new domains by updating the instance preview.

\section{User Study Methodology}\label{sec:user}
The goal of the user study was twofold: to understand the real-world applications and limitations of \reps{}, and to evaluate~\system's usability for discovering systematic failures.
To derive these insights, we conducted semi-structured interviews and think-aloud studies with 10 participants. 
The 10 participants were recruited through university and company email lists. They all had at least three years of ML experience and either published an ML research project or deployed a production AI system.
8 were Ph.D. candidates in a computer science-related discipline (machine learning, human-computer interaction, or computer science), and 2 were developers at different software companies who work with deployed models.
The study lasted up to one hour, and participants were compensated with a \$20 Amazon gift card.

All studies were done remotely over video conferencing and consisted of three primary sections. 
The first part of the study was a semi-structured interview to understand the types of AI systems participants work with and their current process for discovering and fixing systematic failures.
The second part of the study aimed to evaluate the usability and workflow of~\system~by having participants use the system to create and validate hypotheses. 
After a ten-minute introduction and walk-through of the system's major features, we tasked participants with thinking aloud while creating and testing two or three failure hypotheses.
We then provided them with additional images to upload into~\system~and try one of the hypothesis testing strategies.
In the final section, we discussed with participants the potential applications and limitations of using \reps{} in their own models and systems.

Participants used a web-hosted version of~\system~loaded with the reports collected for the domain of eyeglass detection described in Sections~\ref{sec:example}~and~\ref{sec:collecting}.
We focused on eyeglass detection since facial recognition systems are widespread and have received attention for real-world issues~\cite{buolamwini2018gender}.

\section{User Study Results}
Two researchers independently analyzed the results of the user study using thematic analysis and affinity diagramming. 
Additionally, the hypotheses participants generated using~\system~were aggregated in Table~\ref{table:prod-hypos}.
We found that with 10 participants the themes for the applications of \reps{}, usability of~\system, and failures participants discovered (Table~\ref{table:prod-hypos}) converged significantly. 

\subsection{\system~and the Sensemaking Process}

\para{The crowd auditing workflow.} 
All participants found the workflow of~\system~to follow their mental process for discovering and testing systematic errors.
One participant who conducts education research described their model debugging process specifically in terms of sensemaking: ``... we're going to get together and talk and say like, what are our hypotheses about what's really going on here? [...] And then we test the hypothesis by saying: Okay. Here's what we think might be going on. Can we test whether that seems to be true? If we look at thousands of students rather than just this one instance.''

Other participants described similar debugging processes in their own work and identified benefits specific to a crowd auditing process with \reps{}.
One participant who works at a self-driving car company stated that ``This is basically the exact workflow at some level that we kind of do, but it's more of a manual process and it's usually after the fact instead of before the fact.''
Another participant stated that they ``become the crowd and try to simulate the test set'' when validating their models and concluded that our process ``is great for democratizing AI.''

\setlength{\tabcolsep}{10pt}
\renewcommand{\arraystretch}{1.2}
\begin{table}
    \centering
    \caption{\textbf{Participant Hypotheses.} The hypotheses participants generated in the study using~\system. We joined similar hypotheses together and show how many participants reported each one.}\label{table:prod-hypos}
    \begin{tabular}{|l|c|}
        \hline
        Hypothesis & \# of Participants \\
        \hline
        Thin, clear, or no rims	&7 \\
        Dark or tinted lenses	&5 \\ 
        Eyes occluded	&4 \\
        Bad image quality	&2 \\
        Looking sideways	&2 \\
        Eyebrows confused with frames	&1 \\
        Shadow over eyes	&1 \\
        Oddly positioned glasses	&1 \\
        \hline
    \end{tabular}
\end{table}

\para{Exploring failure reports with the report embedding.} 
Eight of the developers came up with their hypotheses from looking at the~\embed, while two had preexisting ideas of potential errors they either looked for in the embedding or used the search functionality to discover. 
While some participants noticed that the embedding contained extraneous concepts, for example, \textit{glasses}, they generally found the embedding useful, with one participant mentioning it was ``very good at finding patterns.''

While all participants eventually found the ~\reps~ useful and understood what they represented, it took two participants a few interactions to fully understand them. 
These participants were initially confused about both where the reports came from and what they represented.
One participant thought that the reports were hypotheses that other developers had created, while another did not know the reports were only for wrongly classified images.
We do not believe this is a severe flaw in the~\system~system or crowd auditing process since, in a deployed setting, the developer would be more involved in defining and collecting the~\reps.
The developer would then have a clear mental model of what the raw data being visualized is.

\para{Creating and validating failure hypotheses.} 
All participants found the~\hypo~interface useful for tracking and testing hypotheses.
There was also significant overlap in the systematic failures participants described, as seen in Table~\ref{table:prod-hypos}, suggesting that~\system~can surface the most prevalent patterns present in a set of~\reps.

An interesting dynamic we discovered during the studies was three participants' uncertainty with how specific to make their hypotheses.
One participant was unsure whether to create a hypothesis for ``face is obstructed'', or the more specific hypothesis of ``hair is covering their face''.
Another participant overcame this uncertainty by beginning with a more general hypothesis and then refining it if, through testing, the general hypothesis was not confirmed.
We intentionally let the developer choose the granularity of hypotheses since, depending on the domain, certain description levels may be more helpful for fixing the discovered issues. 

Another interaction four participants wished for was richer interactions between hypotheses.
In~\system~each hypothesis has its own isolated sets of reports and instances.
Participants at times wished they could drag and drop additional instances into other hypotheses or reuse some of their collected reports.
Richer interactions between hypotheses could create a more seamless experience.

All participants found the additional instance validation useful for testing their hypotheses. 
The modified instance interaction also reflected participants' expectations. 
For example, before being introduced to the system, one participant mentioned that ``counterfactuals seem like a logical choice'' for testing their model.

\subsection{Applications and Limitations of Failure Reports}

\para{Suitable domains.}
Through conversations with our participants, we found a pattern for which AI systems would be best suited for crowd auditing using \reps{}.
Our participants worked with various models, data types, and domains in both industry and research.
We found that the primary factor dictating the usefulness of~\system~was how human-understandable the model's input and output are.
For an end-user to generate a probable explanation, they have to understand what the data and output represent.
For example, financial data or signal processing will generally not work well, as it is often difficult for humans to spot potential issues or edge cases.
This is not the case for domains like pedestrian detection and image captioning, where it is often evident to humans when a model is wrong and what the problem is.  
Many AI systems focus on these human-understandable domains for which end-users can provide viable reports, and over half of our participants worked with such systems.
Failure reports can even be used in more complex domains if domain experts are seeing the model output and can describe potential failures, for example, radiologists and x-rays. 

\para{Model errors in practice.} 
Nine participants had encountered situations in which their models had systematic errors in practice.
For example, one participant developed a system for tracking hands using a video feed and found that the model tended to fail when a hand was at the edge of the screen.
After some investigation, they attributed this issue to a ``center bias'' in the data, which was mostly made up of hands in the middle of the video or picture frame.
Following this example, most participants attributed their system's systematic failures to the data rather than the model.

One of our participants works on a production AI system with multiple chained models.
In this case, it is hard to pinpoint the exact source of the problem, i.e., which model is at fault for the failure, using the testing mechanisms in~\system. 
Despite this, ~\reps~ can still be used to characterize failures the overall system is making.

\section{Experimental Validity of Failure Reports}\label{sec:expirement}

While the user study showed that developers can find and validate various types of systematic failures using~\system, there is an open question if these insights can be used to improve model performance.
To directly test the final stage of our crowd auditing process, Model Improvement, we gathered additional data from the hypotheses participants identified in the user study and retrained the glasses detection model.
We hypothesized that adding data specifically from systematic failure groups will improve a model more than adding the same type of data.
If this hypothesis holds, it suggests images from the blind spots developers found were underrepresented in the data, leading to the failures crowd auditing was able to detect with~\reps~and~\system.

\begin{figure}
    \centering
    \includegraphics[width=\textwidth]{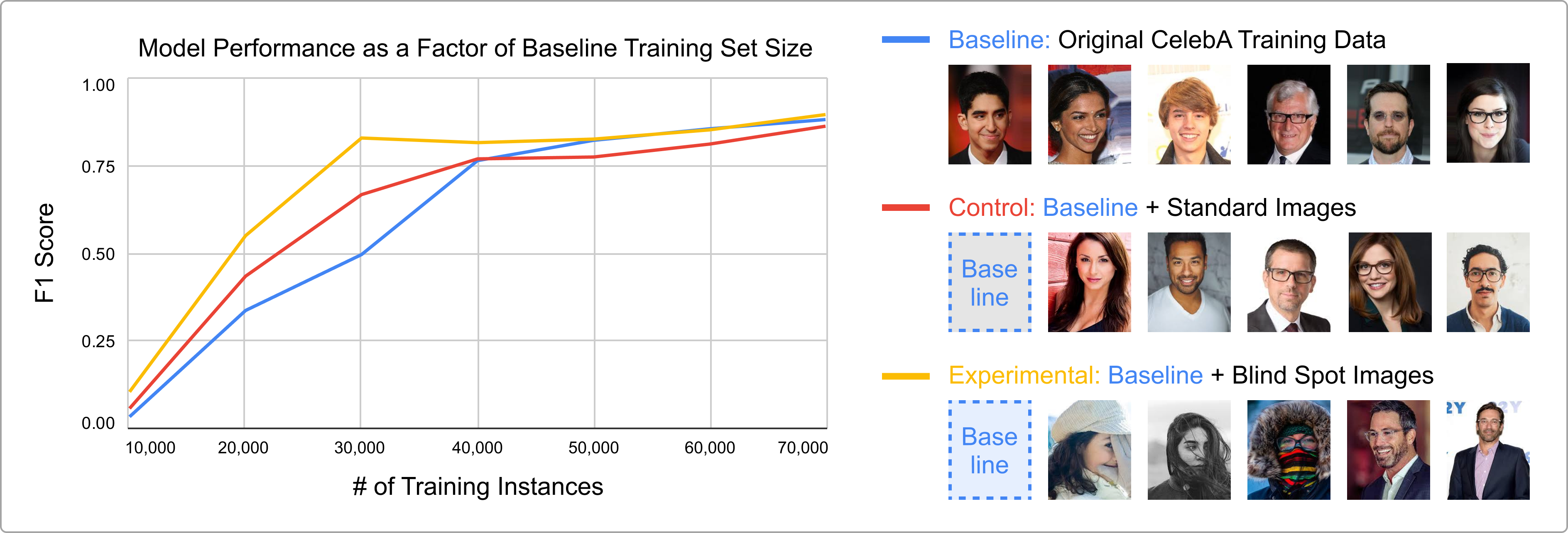}
    \caption{Retraining the eyeglass detection model with data from the blind spots identified with~\system~shows significant improvement in performance over retraining with randomly selected additional data.}
    \label{fig:experiment}
\end{figure}

We collected additional data by using a script to scrape Google Images, limited to permissibly licensed images.
To find particular images, we used search terms like `person wearing glasses' and `person with covered face'.
We then downloaded the images for each search term and removed those that did not fit the description. 
This data gathering process can also be done using the similar image search functionality in \system{}

To isolate the impact of blind spot images on model performance, we used three different conditions.
The baseline condition was a random subset of the original CelebA training data, which has 160,000 headshots of celebrities~\cite{liu2015faceattributes}. 
The control condition consisted of the images from the baseline condition plus 644 additional images: 354 images of people with glasses and 290 of people without glasses. 
The final experimental condition also consisted of the baseline images plus a different set of 673 additional images: 361 images of people with glasses (145 with clear glasses and 215 with covered or occluded faces and glasses) and 312 images of people without glasses and covered or occluded faces.
These additional images represent the most common blind spots discovered by participants in the user study, as seen in Table~\ref{table:prod-hypos}. 
For both the control and experimental condition, we originally collected 750 images each but removed the images that did not fit the search terms.

To measure the impact of additional images on model performance, we retrained every model at different-sized subsets of baseline training images.
When training the baseline model, 600 random images from the training set were added to control for the additional data in the other conditions.
The results can be seen in Figure~\ref{fig:experiment}, with the F1 score used to measure performance on the test set since it gives a more accurate view of performance for imbalanced data.  
We find that at almost every training size, the experimental condition performs better than both the control and baseline conditions, confirming our hypothesis that collecting data from blind spots can better improve model performance.

There are several additional results from this experiment.
First, the control condition performed consistently better than the baseline condition.
We believe that this is likely due to the new images differing from the training data in dimensions like image quality and face positioning that help the model generalize, a phenomenon that has been shown in ImageNet models~\cite{Recht2019}.
Second, we find that the impact of additional images on performance diminishes as the training set size increases.
This is likely due to the number of additional images (\char`\~600) being two orders of magnitude smaller than the original training set (40,000-70,000), minimizing the impact of additional images on training.  

\section{Discussion}

\subsection{Generalization of Failure Reports and Deblinder}\label{sec:captioning}

Our implementation of \system{} represents one example design of how to make \reps{} useful for discovering systematic failures.
Notably, \system~is focused on the setting of one analyst looking at various \reps{} in the domain of images.
Future work could explore methods for using crowdsourcing to synthesize and make sense of reports, further scaling the manual process enabled by~\system, or developing algorithmic techniques for extracting insights. 
This work also focused on text-based failure reports as the primary medium of reporting.
More advanced techniques like data tagging could lead to better methods for clustering similar instances and extracting systematic failures.

The current version of \system{} generalizes to image domains beyond binary classification.
We show this by collecting and visualizing~\reps~for an image captioning model on pictures of animals. 
We used a pre-trained PyTorch captioning model\footnote{\url{https://github.com/yunjey/pytorch-tutorial/tree/master/tutorials/03-advanced/image_captioning}} and a subset of images from the Flickr30K dataset~\cite{Plummer2017}.
In this case, how the model failed is not obvious to end-users, as it might fail to recognize the objects in the image correctly or fail to generate syntactical text.  For this reason, we collected~\reps~for \textit{how} the model failed instead of why in contrast to the eyeglass detection example.
We used the same interface and Amazon Mechanical Turk process as described in Section~\ref{sec:collecting}.  

The resulting~\embed~for image captioning can be seen in Figure~\ref{fig:use-cases}B, and demonstrates some interesting insights.
These include \textit{frisbee} being a common concept; as it turns out, the captioning model describes many standing dogs as holding a frisbee regardless of whether they are holding one or not.
\textit{Running} is also prevalent, as the model describes animals as `running' if they are just standing or upright.
While~\system~was designed to work with any image domain, it is also extensible to include other domains like video, text, and tabular data. 

\subsection{Limitations and Future Work}
Failure reports provide a valuable snapshot into AI systems' real-world systematic errors, but model blind spots are complex with various dimensions.
Extensions to~\system~and future work could further developers' ability to understand and improve their model's real-world performance.

\para{Fixing systematic failures.}
\system{}'s similar image search can be used to collect more training data and improve model performance, but it is not a complete solution.
Future work could explore more effective methods for fixing the detected failures.
Algorithmic techniques like clustering and classification could supplement similar image search and be used to label more training data~\cite{Kim2018a, Liu2020}.
For natural images, this includes pre-trained models for image classification and object detection like those for OpenImages~\cite{Kuznetsova2020}.
Even without more data, recent data programming methods like slice-based learning can be used to improve model performance in areas of high error~\cite{Chen2019a}.

\para{Better testing and tracking.}
While developers appreciated the hypothesis testing techniques, they require nontrivial developer effort and only provide approximations to the ground truth validity of hypotheses.
To scale the testing of hypotheses, crowd-based systems similar to Beat the Machine could be used to task crowd workers with editing or discovering new images~\cite{attenberg2011beat}.
Research into scalable and novel testing techniques could also help improve hypothesis evaluation.
These include more advanced algorithmic methods for finding additional instances or solutions like Generative Adversarial Nets (GANs) that could create synthetic examples to test a blind spot hypothesis~\cite{Radford2016}.
Additional validation methods would also help counter confirmation bias that might come from confounders or biases in individual sources of evidence.

As developers begin to change and improve AI systems, it is important to track their performance over time.
Is the model doing better for some regions of error? Is it regressing for others? Are there new blind spots?
Tracking an AI system's performance 
\system~could be extended to visualize the evolution of failures over time, tracking whether data and model changes have their desired effect.

\para{Complex hypotheses and domain knowledge.}
The~\system~system and the overall process generally matched developers' model debugging processes, but we discovered some limitations and opportunities for improvement in our user study.
\system~was designed to work with discrete, separable blind spots, but we found that they are often much more ambiguously defined.
Some participants wanted to be able to nest and group hypotheses, like, for example, putting the `hat covering face' hypothesis in the more general hypothesis for `face is obstructed'.
This ambiguity could impact later stages of the process like what type of data is collected.
Future work could explore more complex methods for visualizing and organizing hypotheses and supporting evidence.

Another interaction that participants desired was a more direct way to include their domain knowledge and insights into the~\embed~and system.
Richer interactions with the embedding, such as allowing developers to add their own tags and descriptions to ~\reps, could deepen the insights developers can get from the reports.

\para{Better understanding failure reports.}
We have shown for two distinct domains that end-users provide probable~\reps, and experimentally showed for one of the domains that those findings can be used to improve model performance.
While developers can verify blind spots in~\system, future work could explore human descriptions of model errors further, for example, understanding the types of failures people are more likely to detect.
A more complex understanding of the human side of crowd auditing could help developers decide when to deploy the process and guide the development of techniques and systems for encouraging better reporting.

\section{Conclusion}
AI systems are being deployed to a growing number of societally impactful domains.
To improve their performance and understand their potential impacts, developers must know what types of errors their models are making. 
We introduce \n{crowdsourced} \textit{failure reports}, descriptions of how or why a model failed, and show how they can be used to detect and validate systematic failures in AI systems.
To synthesize hundreds or thousands of text reports, we design and implement a visual analytics system,~\system, for aggregating, visualizing, and validating insights from~\reps.
Tightening the loop between model development and real-world evaluation is essential for developing safe and responsible AI systems, and insights from end-users provide a rich new data source for augmenting this process.

\begin{acks}
  This material is based upon work supported by an Amazon grant, a Block Center for Technology and Society grant, a National Science Foundation grant under No. IIS-2040942, and the National Science Foundation Graduate Research Fellowship Program under grant No. DGE-1745016. Any opinions, findings, and conclusions or recommendations expressed in this material are those of the authors and do not necessarily reflect the views of Amazon, the Block Center or the National Science Foundation.
\end{acks}
\bibliographystyle{ACM-Reference-Format}
\bibliography{references,references-static}


\begin{thebibliography}{67}


\ifx \showCODEN    \undefined \def \showCODEN     #1{\unskip}     \fi
\ifx \showDOI      \undefined \def \showDOI       #1{#1}\fi
\ifx \showISBNx    \undefined \def \showISBNx     #1{\unskip}     \fi
\ifx \showISBNxiii \undefined \def \showISBNxiii  #1{\unskip}     \fi
\ifx \showISSN     \undefined \def \showISSN      #1{\unskip}     \fi
\ifx \showLCCN     \undefined \def \showLCCN      #1{\unskip}     \fi
\ifx \shownote     \undefined \def \shownote      #1{#1}          \fi
\ifx \showarticletitle \undefined \def \showarticletitle #1{#1}   \fi
\ifx \showURL      \undefined \def \showURL       {\relax}        \fi
\providecommand\bibfield[2]{#2}
\providecommand\bibinfo[2]{#2}
\providecommand\natexlab[1]{#1}
\providecommand\showeprint[2][]{arXiv:#2}

\bibitem[\protect\citeauthoryear{Ahn and Lin}{Ahn and Lin}{2020}]%
        {Ahn2019}
\bibfield{author}{\bibinfo{person}{Yongsu Ahn} {and} \bibinfo{person}{Yu~Ru
  Lin}.} \bibinfo{year}{2020}\natexlab{}.
\newblock \showarticletitle{{Fairsight: Visual analytics for fairness in
  decision making}}.
\newblock \bibinfo{journal}{\emph{IEEE Transactions on Visualization and
  Computer Graphics}} \bibinfo{volume}{26}, \bibinfo{number}{1}
  (\bibinfo{year}{2020}), \bibinfo{pages}{1086--1095}.
\newblock
\showISSN{19410506}
\urldef\tempurl%
\url{https://doi.org/10.1109/TVCG.2019.2934262}
\showDOI{\tempurl}


\bibitem[\protect\citeauthoryear{Amershi, Begel, Bird, DeLine, Gall, Kamar,
  Nagappan, Nushi, and Zimmermann}{Amershi et~al\mbox{.}}{2019}]%
        {Amershi2019}
\bibfield{author}{\bibinfo{person}{Saleema Amershi}, \bibinfo{person}{Andrew
  Begel}, \bibinfo{person}{Christian Bird}, \bibinfo{person}{Robert DeLine},
  \bibinfo{person}{Harald Gall}, \bibinfo{person}{Ece Kamar},
  \bibinfo{person}{Nachiappan Nagappan}, \bibinfo{person}{Besmira Nushi}, {and}
  \bibinfo{person}{Thomas Zimmermann}.} \bibinfo{year}{2019}\natexlab{}.
\newblock \showarticletitle{{Software Engineering for Machine Learning: A Case
  Study}}.
\newblock \bibinfo{journal}{\emph{Proceedings - 2019 IEEE/ACM 41st
  International Conference on Software Engineering: Software Engineering in
  Practice, ICSE-SEIP 2019}} (\bibinfo{year}{2019}), \bibinfo{pages}{291--300}.
\newblock
\showISBNx{9781728117607}
\urldef\tempurl%
\url{https://doi.org/10.1109/ICSE-SEIP.2019.00042}
\showDOI{\tempurl}


\bibitem[\protect\citeauthoryear{Andr{\'{e}}, Kittur, and Dow}{Andr{\'{e}}
  et~al\mbox{.}}{2014}]%
        {Andre2014}
\bibfield{author}{\bibinfo{person}{Paul Andr{\'{e}}}, \bibinfo{person}{Aniket
  Kittur}, {and} \bibinfo{person}{Steven~P. Dow}.}
  \bibinfo{year}{2014}\natexlab{}.
\newblock \showarticletitle{{Crowd synthesis: Extracting categories and
  clusters from complex data}}.
\newblock \bibinfo{journal}{\emph{Proceedings of the ACM Conference on Computer
  Supported Cooperative Work, CSCW}} (\bibinfo{year}{2014}),
  \bibinfo{pages}{989--998}.
\newblock
\showISBNx{9781450325400}
\urldef\tempurl%
\url{https://doi.org/10.1145/2531602.2531653}
\showDOI{\tempurl}


\bibitem[\protect\citeauthoryear{Attenberg, Ipeirotis, and Provost}{Attenberg
  et~al\mbox{.}}{2011}]%
        {attenberg2011beat}
\bibfield{author}{\bibinfo{person}{Josh~M Attenberg},
  \bibinfo{person}{Pagagiotis~G Ipeirotis}, {and} \bibinfo{person}{Foster
  Provost}.} \bibinfo{year}{2011}\natexlab{}.
\newblock \showarticletitle{Beat the machine: Challenging workers to find the
  unknown unknowns}. In \bibinfo{booktitle}{\emph{Workshops at the Twenty-Fifth
  AAAI Conference on Artificial Intelligence}}.
\newblock


\bibitem[\protect\citeauthoryear{Bansal and Weld}{Bansal and Weld}{2018}]%
        {Bansal2018}
\bibfield{author}{\bibinfo{person}{Gagan Bansal} {and}
  \bibinfo{person}{Daniel~S. Weld}.} \bibinfo{year}{2018}\natexlab{}.
\newblock \showarticletitle{{A coverage-based utility model for identifying
  unknown unknowns}}.
\newblock \bibinfo{journal}{\emph{32nd AAAI Conference on Artificial
  Intelligence, AAAI 2018}} (\bibinfo{year}{2018}),
  \bibinfo{pages}{1463--1470}.
\newblock
\showISBNx{9781577358008}


\bibitem[\protect\citeauthoryear{Barocas and Selbst}{Barocas and
  Selbst}{2018}]%
        {Barocas2016}
\bibfield{author}{\bibinfo{person}{Solon Barocas} {and}
  \bibinfo{person}{Andrew~D Selbst}.} \bibinfo{year}{2018}\natexlab{}.
\newblock \showarticletitle{{Big Data's Disparate Impact}}.
\newblock \bibinfo{journal}{\emph{SSRN Electronic Journal}}
  \bibinfo{volume}{671} (\bibinfo{year}{2018}), \bibinfo{pages}{671--732}.
\newblock
\urldef\tempurl%
\url{https://doi.org/10.2139/ssrn.2477899}
\showDOI{\tempurl}


\bibitem[\protect\citeauthoryear{Bederson, Meyer, and Good}{Bederson
  et~al\mbox{.}}{2000}]%
        {Bederson2000}
\bibfield{author}{\bibinfo{person}{B.~B. Bederson}, \bibinfo{person}{J. Meyer},
  {and} \bibinfo{person}{L. Good}.} \bibinfo{year}{2000}\natexlab{}.
\newblock \showarticletitle{{Jazz: An extensible zoomable user interface
  graphics toolkit in Java}}.
\newblock \bibinfo{journal}{\emph{UIST (User Interface Software and
  Technology): Proceedings of the ACM Symposium}} (\bibinfo{year}{2000}),
  \bibinfo{pages}{171--180}.
\newblock
\urldef\tempurl%
\url{https://doi.org/10.1016/b978-155860915-0/50016-0}
\showDOI{\tempurl}


\bibitem[\protect\citeauthoryear{Bettenburg, Just, Schr\"{o}ter, Weiss,
  Premraj, and Zimmermann}{Bettenburg et~al\mbox{.}}{2008}]%
        {Bettenburg2008what}
\bibfield{author}{\bibinfo{person}{Nicolas Bettenburg}, \bibinfo{person}{Sascha
  Just}, \bibinfo{person}{Adrian Schr\"{o}ter}, \bibinfo{person}{Cathrin
  Weiss}, \bibinfo{person}{Rahul Premraj}, {and} \bibinfo{person}{Thomas
  Zimmermann}.} \bibinfo{year}{2008}\natexlab{}.
\newblock \showarticletitle{What Makes a Good Bug Report?}. In
  \bibinfo{booktitle}{\emph{Proceedings of the 16th ACM SIGSOFT International
  Symposium on Foundations of Software Engineering}} (Atlanta, Georgia)
  \emph{(\bibinfo{series}{SIGSOFT ’08/FSE-16})}.
  \bibinfo{publisher}{Association for Computing Machinery},
  \bibinfo{address}{New York, NY, USA}, \bibinfo{pages}{308–318}.
\newblock
\showISBNx{9781595939951}
\urldef\tempurl%
\url{https://doi.org/10.1145/1453101.1453146}
\showDOI{\tempurl}


\bibitem[\protect\citeauthoryear{Bettenburg, Premraj, Zimmermann, and
  Kim}{Bettenburg et~al\mbox{.}}{[n.d.]}]%
        {bettenburg2008duplicate}
\bibfield{author}{\bibinfo{person}{Nicolas Bettenburg}, \bibinfo{person}{Rahul
  Premraj}, \bibinfo{person}{Thomas Zimmermann}, {and} \bibinfo{person}{Sunghun
  Kim}.} \bibinfo{year}{[n.d.]}\natexlab{}.
\newblock \showarticletitle{Duplicate bug reports considered harmful…
  really?}. In \bibinfo{booktitle}{\emph{2008 IEEE International Conference on
  Software Maintenance}}. IEEE, \bibinfo{pages}{337--345}.
\newblock


\bibitem[\protect\citeauthoryear{Buolamwini and Gebru}{Buolamwini and
  Gebru}{2018}]%
        {buolamwini2018gender}
\bibfield{author}{\bibinfo{person}{Joy Buolamwini} {and}
  \bibinfo{person}{Timnit Gebru}.} \bibinfo{year}{2018}\natexlab{}.
\newblock \showarticletitle{Gender shades: Intersectional accuracy disparities
  in commercial gender classification}. In \bibinfo{booktitle}{\emph{Conference
  on fairness, accountability and transparency}}. \bibinfo{pages}{77--91}.
\newblock


\bibitem[\protect\citeauthoryear{Cabrera, Epperson, Hohman, Kahng, Morgenstern,
  and Chau}{Cabrera et~al\mbox{.}}{2019}]%
        {Cabrera2019}
\bibfield{author}{\bibinfo{person}{Ángel~Alexander Cabrera},
  \bibinfo{person}{Will Epperson}, \bibinfo{person}{Fred Hohman},
  \bibinfo{person}{Minsuk Kahng}, \bibinfo{person}{Jamie Morgenstern}, {and}
  \bibinfo{person}{Duen~Horng Chau}.} \bibinfo{year}{2019}\natexlab{}.
\newblock \showarticletitle{{FairVis: Visual Analytics for Discovering
  Intersectional Bias in Machine Learning}}. In \bibinfo{booktitle}{\emph{2019
  IEEE Conference on Visual Analytics Science and Technology, VAST 2019 -
  Proceedings}}. \bibinfo{pages}{46--56}.
\newblock
\showISBNx{9781728122847}
\urldef\tempurl%
\url{https://doi.org/10.1109/VAST47406.2019.8986948}
\showDOI{\tempurl}


\bibitem[\protect\citeauthoryear{Cabrera, Hohman, Lin, and Chau}{Cabrera
  et~al\mbox{.}}{2018}]%
        {Cabrera2018}
\bibfield{author}{\bibinfo{person}{Ángel~Alexander Cabrera},
  \bibinfo{person}{Fred Hohman}, \bibinfo{person}{Jason Lin}, {and}
  \bibinfo{person}{Duen~Horng Chau}.} \bibinfo{year}{2018}\natexlab{}.
\newblock \showarticletitle{{Interactive Classification for Deep Learning
  Interpretation}}.
\newblock  (\bibinfo{year}{2018}), \bibinfo{pages}{1--5}.
\newblock
\urldef\tempurl%
\url{http://arxiv.org/abs/1806.05660}
\showURL{%
\tempurl}


\bibitem[\protect\citeauthoryear{Chang, Kittur, and Hahn}{Chang
  et~al\mbox{.}}{2016}]%
        {Chang2016}
\bibfield{author}{\bibinfo{person}{Joseph~Chee Chang}, \bibinfo{person}{Aniket
  Kittur}, {and} \bibinfo{person}{Nathan Hahn}.}
  \bibinfo{year}{2016}\natexlab{}.
\newblock \showarticletitle{{Alloy: Clustering with crowds and computation}}.
\newblock \bibinfo{journal}{\emph{Conference on Human Factors in Computing
  Systems - Proceedings}} (\bibinfo{year}{2016}), \bibinfo{pages}{3180--3191}.
\newblock
\showISBNx{9781450333627}
\urldef\tempurl%
\url{https://doi.org/10.1145/2858036.2858411}
\showDOI{\tempurl}


\bibitem[\protect\citeauthoryear{Chau, Kittur, Hong, and Faloutsos}{Chau
  et~al\mbox{.}}{2011}]%
        {Chau2011}
\bibfield{author}{\bibinfo{person}{Duen~Horng Chau}, \bibinfo{person}{Aniket
  Kittur}, \bibinfo{person}{Jason~I. Hong}, {and} \bibinfo{person}{Christos
  Faloutsos}.} \bibinfo{year}{2011}\natexlab{}.
\newblock \showarticletitle{{Apolo: Making sense of large network data by
  combining rich user interaction and machine learning}}.
\newblock \bibinfo{journal}{\emph{Conference on Human Factors in Computing
  Systems - Proceedings}} (\bibinfo{year}{2011}), \bibinfo{pages}{167--176}.
\newblock
\showISBNx{9781450302289}
\urldef\tempurl%
\url{https://doi.org/10.1145/1978942.1978967}
\showDOI{\tempurl}


\bibitem[\protect\citeauthoryear{Chen, Suh, Verwey, Ramos, Drucker, and
  Simard}{Chen et~al\mbox{.}}{2018}]%
        {Chen2018}
\bibfield{author}{\bibinfo{person}{Nan~Chen Chen}, \bibinfo{person}{Jina Suh},
  \bibinfo{person}{Johan Verwey}, \bibinfo{person}{Gonzalo Ramos},
  \bibinfo{person}{Steven Drucker}, {and} \bibinfo{person}{Patrice Simard}.}
  \bibinfo{year}{2018}\natexlab{}.
\newblock \showarticletitle{{Anchorviz: Facilitating classifier error discovery
  through interactive semantic data exploration}}.
\newblock \bibinfo{journal}{\emph{International Conference on Intelligent User
  Interfaces, Proceedings IUI}} (\bibinfo{year}{2018}),
  \bibinfo{pages}{269--280}.
\newblock
\showISBNx{9781450349451}
\urldef\tempurl%
\url{https://doi.org/10.1145/3172944.3172950}
\showDOI{\tempurl}


\bibitem[\protect\citeauthoryear{Chen, Wu, Weng, Ratner, and R{\'{e}}}{Chen
  et~al\mbox{.}}{2019}]%
        {Chen2019a}
\bibfield{author}{\bibinfo{person}{Vincent~S. Chen}, \bibinfo{person}{Sen Wu},
  \bibinfo{person}{Zhenzhen Weng}, \bibinfo{person}{Alexander Ratner}, {and}
  \bibinfo{person}{Christopher R{\'{e}}}.} \bibinfo{year}{2019}\natexlab{}.
\newblock \showarticletitle{{Slice-based Learning: A Programming Model for
  Residual Learning in Critical Data Slices}}.
\newblock  \bibinfo{number}{NeurIPS} (\bibinfo{year}{2019}).
\newblock
\showISSN{1049-5258}
\urldef\tempurl%
\url{http://arxiv.org/abs/1909.06349}
\showURL{%
\tempurl}


\bibitem[\protect\citeauthoryear{Cheng and Bernstein}{Cheng and
  Bernstein}{2015}]%
        {Cheng2015}
\bibfield{author}{\bibinfo{person}{Justin Cheng} {and}
  \bibinfo{person}{Michael~S. Bernstein}.} \bibinfo{year}{2015}\natexlab{}.
\newblock \showarticletitle{{Flock: Hybrid crowd-machine learning
  classifiers}}.
\newblock \bibinfo{journal}{\emph{CSCW 2015 - Proceedings of the 2015 ACM
  International Conference on Computer-Supported Cooperative Work and Social
  Computing}} (\bibinfo{year}{2015}), \bibinfo{pages}{600--611}.
\newblock
\showISBNx{9781450329224}
\urldef\tempurl%
\url{https://doi.org/10.1145/2675133.2675214}
\showDOI{\tempurl}


\bibitem[\protect\citeauthoryear{Chung, Kraska, Polyzotis, Tae, and
  Whang}{Chung et~al\mbox{.}}{2019}]%
        {Chung2019}
\bibfield{author}{\bibinfo{person}{Yeounoh Chung}, \bibinfo{person}{Tim
  Kraska}, \bibinfo{person}{Neoklis Polyzotis}, \bibinfo{person}{Ki~Hyun Tae},
  {and} \bibinfo{person}{Steven~Euijong Whang}.}
  \bibinfo{year}{2019}\natexlab{}.
\newblock \showarticletitle{{Slice finder: Automated data slicing for model
  validation}}.
\newblock \bibinfo{journal}{\emph{Proceedings - International Conference on
  Data Engineering}}  \bibinfo{volume}{2019-April} (\bibinfo{year}{2019}),
  \bibinfo{pages}{1550--1553}.
\newblock
\showISBNx{9781538674741}
\showISSN{10844627}
\urldef\tempurl%
\url{https://doi.org/10.1109/ICDE.2019.00139}
\showDOI{\tempurl}


\bibitem[\protect\citeauthoryear{Felix, Franconeri, and Bertini}{Felix
  et~al\mbox{.}}{2018}]%
        {Felix2018}
\bibfield{author}{\bibinfo{person}{Cristian Felix}, \bibinfo{person}{Steven
  Franconeri}, {and} \bibinfo{person}{Enrico Bertini}.}
  \bibinfo{year}{2018}\natexlab{}.
\newblock \showarticletitle{{Taking Word Clouds Apart: An Empirical
  Investigation of the Design Space for Keyword Summaries}}.
\newblock \bibinfo{journal}{\emph{IEEE Transactions on Visualization and
  Computer Graphics}} \bibinfo{volume}{24}, \bibinfo{number}{1}
  (\bibinfo{year}{2018}), \bibinfo{pages}{657--666}.
\newblock
\showISSN{10772626}
\urldef\tempurl%
\url{https://doi.org/10.1109/TVCG.2017.2746018}
\showDOI{\tempurl}


\bibitem[\protect\citeauthoryear{Fisher, Counts, and Kittur}{Fisher
  et~al\mbox{.}}{2012}]%
        {Fisher2012}
\bibfield{author}{\bibinfo{person}{Kristie Fisher}, \bibinfo{person}{Scott
  Counts}, {and} \bibinfo{person}{Aniket Kittur}.}
  \bibinfo{year}{2012}\natexlab{}.
\newblock \showarticletitle{{Distributed sensemaking: Improving sensemaking by
  leveraging the efforts of previous users}}.
\newblock \bibinfo{journal}{\emph{Conference on Human Factors in Computing
  Systems - Proceedings}} (\bibinfo{year}{2012}), \bibinfo{pages}{247--256}.
\newblock
\showISBNx{9781450310154}
\urldef\tempurl%
\url{https://doi.org/10.1145/2207676.2207711}
\showDOI{\tempurl}


\bibitem[\protect\citeauthoryear{Foong, Gergle, and Gerber}{Foong
  et~al\mbox{.}}{2017}]%
        {Foong2017a}
\bibfield{author}{\bibinfo{person}{Eureka Foong}, \bibinfo{person}{Darren
  Gergle}, {and} \bibinfo{person}{Elizabeth~M. Gerber}.}
  \bibinfo{year}{2017}\natexlab{}.
\newblock \showarticletitle{{Novice and expert sensemaking of crowdsourced
  feedback}}.
\newblock \bibinfo{journal}{\emph{Proceedings of the ACM on Human-Computer
  Interaction}} \bibinfo{volume}{1}, \bibinfo{number}{CSCW}
  (\bibinfo{year}{2017}), \bibinfo{pages}{1--18}.
\newblock
\showISSN{25730142}
\urldef\tempurl%
\url{https://doi.org/10.1145/3134680}
\showDOI{\tempurl}


\bibitem[\protect\citeauthoryear{G{\"{o}}rg, Liu, Kihm, Choo, Park, and
  Stasko}{G{\"{o}}rg et~al\mbox{.}}{2013}]%
        {Gorg2013}
\bibfield{author}{\bibinfo{person}{Carsten G{\"{o}}rg},
  \bibinfo{person}{Zhicheng Liu}, \bibinfo{person}{Jaeyeon Kihm},
  \bibinfo{person}{Jaegul Choo}, \bibinfo{person}{Haesun Park}, {and}
  \bibinfo{person}{John Stasko}.} \bibinfo{year}{2013}\natexlab{}.
\newblock \showarticletitle{{Combining computational analyses and interactive
  visualization for document exploration and sensemaking in jigsaw}}.
\newblock \bibinfo{journal}{\emph{IEEE Transactions on Visualization and
  Computer Graphics}} \bibinfo{volume}{19}, \bibinfo{number}{10}
  (\bibinfo{year}{2013}), \bibinfo{pages}{1646--1663}.
\newblock
\showISSN{10772626}
\urldef\tempurl%
\url{https://doi.org/10.1109/TVCG.2012.324}
\showDOI{\tempurl}


\bibitem[\protect\citeauthoryear{Goyal}{Goyal}{2015}]%
        {Goyal2015}
\bibfield{author}{\bibinfo{person}{Nitesh Goyal}.}
  \bibinfo{year}{2015}\natexlab{}.
\newblock \showarticletitle{{Designing for Collaborative Sensemaking: Using
  Expert {\&} Non-Expert Crowd}}.
\newblock  (\bibinfo{year}{2015}).
\newblock
\urldef\tempurl%
\url{http://arxiv.org/abs/1511.06053}
\showURL{%
\tempurl}


\bibitem[\protect\citeauthoryear{Heo, Roh, Hwang, Lee, and Whang}{Heo
  et~al\mbox{.}}{2020}]%
        {Heo2020}
\bibfield{author}{\bibinfo{person}{Geon Heo}, \bibinfo{person}{Yuji Roh},
  \bibinfo{person}{Seonghyeon Hwang}, \bibinfo{person}{Dayun Lee}, {and}
  \bibinfo{person}{Steven~Euijong Whang}.} \bibinfo{year}{2020}\natexlab{}.
\newblock \showarticletitle{Inspector Gadget: A Data Programming-Based Labeling
  System for Industrial Images}.
\newblock \bibinfo{journal}{\emph{Proc. VLDB Endow.}} \bibinfo{volume}{14},
  \bibinfo{number}{1} (\bibinfo{date}{Sept.} \bibinfo{year}{2020}),
  \bibinfo{pages}{28–36}.
\newblock
\showISSN{2150-8097}
\urldef\tempurl%
\url{https://doi.org/10.14778/3421424.3421429}
\showDOI{\tempurl}


\bibitem[\protect\citeauthoryear{Holstein, Vaughan, Daum{\'{e}}, Dud{\'{i}}k,
  and Wallach}{Holstein et~al\mbox{.}}{2019}]%
        {Holstein2019}
\bibfield{author}{\bibinfo{person}{Kenneth Holstein},
  \bibinfo{person}{Jennifer~Wortman Vaughan}, \bibinfo{person}{Hal
  Daum{\'{e}}}, \bibinfo{person}{Miroslav Dud{\'{i}}k}, {and}
  \bibinfo{person}{Hanna Wallach}.} \bibinfo{year}{2019}\natexlab{}.
\newblock \showarticletitle{{Improving fairness in machine learning systems:
  What do industry practitioners need?}}
\newblock \bibinfo{journal}{\emph{Conference on Human Factors in Computing
  Systems - Proceedings}} (\bibinfo{year}{2019}), \bibinfo{pages}{1--16}.
\newblock
\showISBNx{9781450359702}
\urldef\tempurl%
\url{https://doi.org/10.1145/3290605.3300830}
\showDOI{\tempurl}


\bibitem[\protect\citeauthoryear{Ipeirotis}{Ipeirotis}{[n.d.]}]%
        {Ipeirotis}
\bibfield{author}{\bibinfo{person}{Panos Ipeirotis}.}
  \bibinfo{year}{[n.d.]}\natexlab{}.
\newblock \showarticletitle{{Demographics of mechanical Turk CeDER-10-01.pdf}}.
\newblock  (\bibinfo{year}{[n.\,d.]}).
\newblock
\urldef\tempurl%
\url{http://archive.nyu.edu/fda/bitstream/2451/29585/2/CeDER-10-01.pdf}
\showURL{%
\tempurl}


\bibitem[\protect\citeauthoryear{Jiang, Li, Ren, Xuan, and Jin}{Jiang
  et~al\mbox{.}}{2018}]%
        {Jiang2018a}
\bibfield{author}{\bibinfo{person}{He Jiang}, \bibinfo{person}{Xiaochen Li},
  \bibinfo{person}{Zhilei Ren}, \bibinfo{person}{Jifeng Xuan}, {and}
  \bibinfo{person}{Zhi Jin}.} \bibinfo{year}{2018}\natexlab{}.
\newblock \showarticletitle{{Toward Better Summarizing Bug Reports With
  Crowdsourcing Elicited Attributes}}.
\newblock \bibinfo{journal}{\emph{IEEE Transactions on Reliability}}
  \bibinfo{volume}{PP} (\bibinfo{year}{2018}), \bibinfo{pages}{1--21}.
\newblock
\urldef\tempurl%
\url{https://doi.org/10.1109/TR.2018.2873427}
\showDOI{\tempurl}


\bibitem[\protect\citeauthoryear{Jindal, Member, and Kaur}{Jindal
  et~al\mbox{.}}{2020}]%
        {Jindal2020}
\bibfield{author}{\bibinfo{person}{Shubhra~Goyal Jindal},
  \bibinfo{person}{Student Member}, {and} \bibinfo{person}{Arvinder Kaur}.}
  \bibinfo{year}{2020}\natexlab{}.
\newblock \showarticletitle{{Automatic Keyword and Sentence-Based Text
  Summarization for Software Bug Reports}}.
\newblock \bibinfo{journal}{\emph{IEEE Access}}  \bibinfo{volume}{8}
  (\bibinfo{year}{2020}), \bibinfo{pages}{65352--65370}.
\newblock
\urldef\tempurl%
\url{https://doi.org/10.1109/ACCESS.2020.2985222}
\showDOI{\tempurl}


\bibitem[\protect\citeauthoryear{Kahng, Fang, and Chau}{Kahng
  et~al\mbox{.}}{2016}]%
        {Kahng2016}
\bibfield{author}{\bibinfo{person}{Minsuk Kahng}, \bibinfo{person}{Dezhi Fang},
  {and} \bibinfo{person}{Duen~Horng Chau}.} \bibinfo{year}{2016}\natexlab{}.
\newblock \showarticletitle{{Visual exploration of machine learning results
  using data cube analysis}}.
\newblock \bibinfo{journal}{\emph{HILDA 2016 - Proceedings of the Workshop on
  Human-In-the-Loop Data Analytics}} (\bibinfo{year}{2016}).
\newblock
\showISBNx{9781450342070}
\urldef\tempurl%
\url{https://doi.org/10.1145/2939502.2939503}
\showDOI{\tempurl}


\bibitem[\protect\citeauthoryear{Kim, Wattenberg, Gilmer, Cai, Wexler, Viegas,
  and Sayres}{Kim et~al\mbox{.}}{2018a}]%
        {Kim2018a}
\bibfield{author}{\bibinfo{person}{Been Kim}, \bibinfo{person}{Martin
  Wattenberg}, \bibinfo{person}{Justin Gilmer}, \bibinfo{person}{Carrie Cai},
  \bibinfo{person}{James Wexler}, \bibinfo{person}{Fernanda Viegas}, {and}
  \bibinfo{person}{Rory Sayres}.} \bibinfo{year}{2018}\natexlab{a}.
\newblock \showarticletitle{{Interpretability beyond feature attribution:
  Quantitative Testing with Concept Activation Vectors (TCAV)}}.
\newblock \bibinfo{journal}{\emph{35th International Conference on Machine
  Learning, ICML 2018}}  \bibinfo{volume}{6} (\bibinfo{year}{2018}),
  \bibinfo{pages}{4186--4195}.
\newblock
\showISBNx{9781510867963}


\bibitem[\protect\citeauthoryear{Kim, Zimmermann, Deline, and Begel}{Kim
  et~al\mbox{.}}{2018b}]%
        {Kim2018}
\bibfield{author}{\bibinfo{person}{Miryung Kim}, \bibinfo{person}{Thomas
  Zimmermann}, \bibinfo{person}{Robert Deline}, {and} \bibinfo{person}{Andrew
  Begel}.} \bibinfo{year}{2018}\natexlab{b}.
\newblock \showarticletitle{{Data scientists in software teams: State of the
  art and challenges}}.
\newblock \bibinfo{journal}{\emph{IEEE Transactions on Software Engineering}}
  \bibinfo{volume}{44}, \bibinfo{number}{11} (\bibinfo{year}{2018}),
  \bibinfo{pages}{1024--1038}.
\newblock
\showISSN{19393520}
\urldef\tempurl%
\url{https://doi.org/10.1109/TSE.2017.2754374}
\showDOI{\tempurl}


\bibitem[\protect\citeauthoryear{Kittur, Peters, Diriye, and Bove}{Kittur
  et~al\mbox{.}}{2014}]%
        {Kittur2014}
\bibfield{author}{\bibinfo{person}{Aniket Kittur}, \bibinfo{person}{Andrew~M.
  Peters}, \bibinfo{person}{Abdigani Diriye}, {and} \bibinfo{person}{Michael~R.
  Bove}.} \bibinfo{year}{2014}\natexlab{}.
\newblock \showarticletitle{{Standing on the schemas of giants: Socially
  augmented information foraging}}.
\newblock \bibinfo{journal}{\emph{Proceedings of the ACM Conference on Computer
  Supported Cooperative Work, CSCW}} (\bibinfo{year}{2014}),
  \bibinfo{pages}{999--1010}.
\newblock
\showISBNx{9781450325400}
\urldef\tempurl%
\url{https://doi.org/10.1145/2531602.2531644}
\showDOI{\tempurl}


\bibitem[\protect\citeauthoryear{Kittur, Peters, Diriye, Telang, and
  Bove}{Kittur et~al\mbox{.}}{2013}]%
        {Kittur2013}
\bibfield{author}{\bibinfo{person}{Aniket Kittur}, \bibinfo{person}{Andrew~M.
  Peters}, \bibinfo{person}{Abdigani Diriye}, \bibinfo{person}{Trupti Telang},
  {and} \bibinfo{person}{Michael~R. Bove}.} \bibinfo{year}{2013}\natexlab{}.
\newblock \showarticletitle{{Costs and benefits of structured information
  foraging}}.
\newblock \bibinfo{journal}{\emph{Conference on Human Factors in Computing
  Systems - Proceedings}} (\bibinfo{year}{2013}), \bibinfo{pages}{2989--2998}.
\newblock
\showISBNx{9781450318990}
\urldef\tempurl%
\url{https://doi.org/10.1145/2470654.2481415}
\showDOI{\tempurl}


\bibitem[\protect\citeauthoryear{Kucher and Kerren}{Kucher and Kerren}{2015}]%
        {Kucher2015}
\bibfield{author}{\bibinfo{person}{Kostiantyn Kucher} {and}
  \bibinfo{person}{Andreas Kerren}.} \bibinfo{year}{2015}\natexlab{}.
\newblock \showarticletitle{{Text visualization techniques: Taxonomy, visual
  survey, and community insights}}.
\newblock \bibinfo{journal}{\emph{IEEE Pacific Visualization Symposium}}
  \bibinfo{volume}{2015-July} (\bibinfo{year}{2015}),
  \bibinfo{pages}{117--121}.
\newblock
\showISBNx{9781467368797}
\showISSN{21658773}
\urldef\tempurl%
\url{https://doi.org/10.1109/PACIFICVIS.2015.7156366}
\showDOI{\tempurl}


\bibitem[\protect\citeauthoryear{Kuznetsova, Rom, Alldrin, Uijlings, Krasin,
  Pont-Tuset, Kamali, Popov, Malloci, Kolesnikov, Duerig, and
  Ferrari}{Kuznetsova et~al\mbox{.}}{2020}]%
        {Kuznetsova2020}
\bibfield{author}{\bibinfo{person}{Alina Kuznetsova}, \bibinfo{person}{Hassan
  Rom}, \bibinfo{person}{Neil Alldrin}, \bibinfo{person}{Jasper Uijlings},
  \bibinfo{person}{Ivan Krasin}, \bibinfo{person}{Jordi Pont-Tuset},
  \bibinfo{person}{Shahab Kamali}, \bibinfo{person}{Stefan Popov},
  \bibinfo{person}{Matteo Malloci}, \bibinfo{person}{Alexander Kolesnikov},
  \bibinfo{person}{Tom Duerig}, {and} \bibinfo{person}{Vittorio Ferrari}.}
  \bibinfo{year}{2020}\natexlab{}.
\newblock \showarticletitle{{The Open Images Dataset V4: Unified Image
  Classification, Object Detection, and Visual Relationship Detection at
  Scale}}.
\newblock \bibinfo{journal}{\emph{International Journal of Computer Vision}}
  \bibinfo{volume}{128}, \bibinfo{number}{7} (\bibinfo{year}{2020}),
  \bibinfo{pages}{1956--1981}.
\newblock
\showISSN{15731405}
\urldef\tempurl%
\url{https://doi.org/10.1007/s11263-020-01316-z}
\showDOI{\tempurl}


\bibitem[\protect\citeauthoryear{Lakkaraju, Kamar, Caruana, and
  Horvitz}{Lakkaraju et~al\mbox{.}}{2017}]%
        {Lakkaraju2017}
\bibfield{author}{\bibinfo{person}{Himabindu Lakkaraju}, \bibinfo{person}{Ece
  Kamar}, \bibinfo{person}{Rich Caruana}, {and} \bibinfo{person}{Eric
  Horvitz}.} \bibinfo{year}{2017}\natexlab{}.
\newblock \showarticletitle{{Identifying unknown unknowns in the open world:
  Representations and policies for guided exploration}}.
\newblock \bibinfo{journal}{\emph{31st AAAI Conference on Artificial
  Intelligence, AAAI 2017}} \bibinfo{number}{Settles 2009}
  (\bibinfo{year}{2017}), \bibinfo{pages}{2124--2132}.
\newblock


\bibitem[\protect\citeauthoryear{Liu, Guerra, Fung, Matute, Kamar, and
  Lasecki}{Liu et~al\mbox{.}}{2020}]%
        {Liu2020}
\bibfield{author}{\bibinfo{person}{Anthony Liu}, \bibinfo{person}{Santiago
  Guerra}, \bibinfo{person}{Isaac Fung}, \bibinfo{person}{Gabriel Matute},
  \bibinfo{person}{Ece Kamar}, {and} \bibinfo{person}{Walter Lasecki}.}
  \bibinfo{year}{2020}\natexlab{}.
\newblock \showarticletitle{{Towards Hybrid Human-AI Workflows for Unknown
  Unknown Detection}}.
\newblock \bibinfo{journal}{\emph{The Web Conference 2020 - Proceedings of the
  World Wide Web Conference, WWW 2020}}  \bibinfo{volume}{2020}
  (\bibinfo{year}{2020}), \bibinfo{pages}{2432--2442}.
\newblock
\showISBNx{9781450370233}
\urldef\tempurl%
\url{https://doi.org/10.1145/3366423.3380306}
\showDOI{\tempurl}


\bibitem[\protect\citeauthoryear{Liu, Luo, Wang, and Tang}{Liu
  et~al\mbox{.}}{2015}]%
        {liu2015faceattributes}
\bibfield{author}{\bibinfo{person}{Ziwei Liu}, \bibinfo{person}{Ping Luo},
  \bibinfo{person}{Xiaogang Wang}, {and} \bibinfo{person}{Xiaoou Tang}.}
  \bibinfo{year}{2015}\natexlab{}.
\newblock \showarticletitle{Deep Learning Face Attributes in the Wild}. In
  \bibinfo{booktitle}{\emph{Proceedings of International Conference on Computer
  Vision (ICCV)}}.
\newblock


\bibitem[\protect\citeauthoryear{Mandel, Best, Tanaka, Temple, Haili, Carter,
  Schlechtinger, and Szeto}{Mandel et~al\mbox{.}}{2020}]%
        {Mandel2020}
\bibfield{author}{\bibinfo{person}{Travis Mandel}, \bibinfo{person}{Jahnu
  Best}, \bibinfo{person}{Randall~H. Tanaka}, \bibinfo{person}{Hiram Temple},
  \bibinfo{person}{Chansen Haili}, \bibinfo{person}{Sebastian~J. Carter},
  \bibinfo{person}{Kayla Schlechtinger}, {and} \bibinfo{person}{Roy Szeto}.}
  \bibinfo{year}{2020}\natexlab{}.
\newblock \showarticletitle{{Using the Crowd to Prevent Harmful AI Behavior}}.
\newblock \bibinfo{journal}{\emph{Proceedings of the ACM on Human-Computer
  Interaction}} \bibinfo{volume}{4}, \bibinfo{number}{CSCW2}
  (\bibinfo{year}{2020}).
\newblock
\showISSN{25730142}
\urldef\tempurl%
\url{https://doi.org/10.1145/3415168}
\showDOI{\tempurl}


\bibitem[\protect\citeauthoryear{McInnes, Healy, Saul, and Grossberger}{McInnes
  et~al\mbox{.}}{2018}]%
        {mcinnes2018umap-software}
\bibfield{author}{\bibinfo{person}{Leland McInnes}, \bibinfo{person}{John
  Healy}, \bibinfo{person}{Nathaniel Saul}, {and} \bibinfo{person}{Lukas
  Grossberger}.} \bibinfo{year}{2018}\natexlab{}.
\newblock \showarticletitle{UMAP: Uniform Manifold Approximation and
  Projection}.
\newblock \bibinfo{journal}{\emph{The Journal of Open Source Software}}
  \bibinfo{volume}{3}, \bibinfo{number}{29} (\bibinfo{year}{2018}),
  \bibinfo{pages}{861}.
\newblock


\bibitem[\protect\citeauthoryear{Moreno-Torres, Raeder, Alaiz-Rodr{\'{i}}guez,
  Chawla, and Herrera}{Moreno-Torres et~al\mbox{.}}{2012}]%
        {Moreno-Torres2012}
\bibfield{author}{\bibinfo{person}{Jose~G. Moreno-Torres},
  \bibinfo{person}{Troy Raeder}, \bibinfo{person}{Rocío
  Alaiz-Rodr{\'{i}}guez}, \bibinfo{person}{Nitesh~V. Chawla}, {and}
  \bibinfo{person}{Francisco Herrera}.} \bibinfo{year}{2012}\natexlab{}.
\newblock \showarticletitle{{A unifying view on dataset shift in
  classification}}.
\newblock \bibinfo{journal}{\emph{Pattern Recognition}} \bibinfo{volume}{45},
  \bibinfo{number}{1} (\bibinfo{year}{2012}), \bibinfo{pages}{521--530}.
\newblock
\showISSN{00313203}
\urldef\tempurl%
\url{https://doi.org/10.1016/j.patcog.2011.06.019}
\showDOI{\tempurl}


\bibitem[\protect\citeauthoryear{Nushi, Kamar, and Horvitz}{Nushi
  et~al\mbox{.}}{2018}]%
        {Nushi2018}
\bibfield{author}{\bibinfo{person}{Besmira Nushi}, \bibinfo{person}{Ece Kamar},
  {and} \bibinfo{person}{Eric Horvitz}.} \bibinfo{year}{2018}\natexlab{}.
\newblock \showarticletitle{{Towards Accountable AI: Hybrid Human-Machine
  Analyses for Characterizing System Failure}}.
\newblock \bibinfo{journal}{\emph{HCOMP}} (\bibinfo{year}{2018}).
\newblock
\urldef\tempurl%
\url{http://arxiv.org/abs/1809.07424}
\showURL{%
\tempurl}


\bibitem[\protect\citeauthoryear{Pennington, Socher, and Manning}{Pennington
  et~al\mbox{.}}{2014}]%
        {pennington2014glove}
\bibfield{author}{\bibinfo{person}{Jeffrey Pennington},
  \bibinfo{person}{Richard Socher}, {and} \bibinfo{person}{Christopher~D
  Manning}.} \bibinfo{year}{2014}\natexlab{}.
\newblock \showarticletitle{Glove: Global vectors for word representation}. In
  \bibinfo{booktitle}{\emph{Proceedings of the 2014 conference on empirical
  methods in natural language processing (EMNLP)}}.
  \bibinfo{pages}{1532--1543}.
\newblock


\bibitem[\protect\citeauthoryear{Pirolli and Card}{Pirolli and Card}{1999}]%
        {Pirolli1999}
\bibfield{author}{\bibinfo{person}{Peter Pirolli} {and} \bibinfo{person}{Stuart
  Card}.} \bibinfo{year}{1999}\natexlab{}.
\newblock \showarticletitle{{Information foraging}}.
\newblock \bibinfo{journal}{\emph{Psychological Review}} \bibinfo{volume}{106},
  \bibinfo{number}{4} (\bibinfo{year}{1999}), \bibinfo{pages}{643--675}.
\newblock
\showISSN{0033295X}
\urldef\tempurl%
\url{https://doi.org/10.1037/0033-295X.106.4.643}
\showDOI{\tempurl}


\bibitem[\protect\citeauthoryear{Pirolli and Card}{Pirolli and Card}{2005}]%
        {Pirolli2005}
\bibfield{author}{\bibinfo{person}{Peter Pirolli} {and} \bibinfo{person}{Stuart
  Card}.} \bibinfo{year}{2005}\natexlab{}.
\newblock \showarticletitle{{The sensemaking process and leverage points for
  analyst technology as identified through cognitive task analysis}}.
\newblock \bibinfo{journal}{\emph{Proceedings of International Conference on
  Intelligence Analysis}} \bibinfo{volume}{2005}, \bibinfo{number}{January}
  (\bibinfo{year}{2005}), \bibinfo{pages}{2–4}.
\newblock
\showISBNx{0029-7844 (Print)}
\showISSN{00219355}
\urldef\tempurl%
\url{https://doi.org/10.1007/s13398-014-0173-7.2}
\showDOI{\tempurl}
\showeprint[arxiv]{9809069v1}~[gr-qc]


\bibitem[\protect\citeauthoryear{Plummer, Wang, Cervantes, Caicedo,
  Hockenmaier, and Lazebnik}{Plummer et~al\mbox{.}}{2017}]%
        {Plummer2017}
\bibfield{author}{\bibinfo{person}{Bryan~A. Plummer}, \bibinfo{person}{Liwei
  Wang}, \bibinfo{person}{Chris~M. Cervantes}, \bibinfo{person}{Juan~C.
  Caicedo}, \bibinfo{person}{Julia Hockenmaier}, {and}
  \bibinfo{person}{Svetlana Lazebnik}.} \bibinfo{year}{2017}\natexlab{}.
\newblock \showarticletitle{{Flickr30k Entities: Collecting Region-to-Phrase
  Correspondences for Richer Image-to-Sentence Models}}.
\newblock \bibinfo{journal}{\emph{International Journal of Computer Vision}}
  \bibinfo{volume}{123}, \bibinfo{number}{1} (\bibinfo{year}{2017}),
  \bibinfo{pages}{74--93}.
\newblock
\showISSN{15731405}
\urldef\tempurl%
\url{https://doi.org/10.1007/s11263-016-0965-7}
\showDOI{\tempurl}


\bibitem[\protect\citeauthoryear{Radford, Metz, and Chintala}{Radford
  et~al\mbox{.}}{2016}]%
        {Radford2016}
\bibfield{author}{\bibinfo{person}{Alec Radford}, \bibinfo{person}{Luke Metz},
  {and} \bibinfo{person}{Soumith Chintala}.} \bibinfo{year}{2016}\natexlab{}.
\newblock \showarticletitle{{Unsupervised representation learning with deep
  convolutional generative adversarial networks}}.
\newblock \bibinfo{journal}{\emph{4th International Conference on Learning
  Representations, ICLR 2016 - Conference Track Proceedings}}
  (\bibinfo{year}{2016}), \bibinfo{pages}{1--16}.
\newblock


\bibitem[\protect\citeauthoryear{Rahwan, Cebrian, Obradovich, Bongard,
  Bonnefon, Breazeal, Crandall, Christakis, Couzin, Jackson, Jennings, Kamar,
  Kloumann, Larochelle, Lazer, McElreath, Mislove, Parkes, Pentland, Roberts,
  Shariff, Tenenbaum, and Wellman}{Rahwan et~al\mbox{.}}{2019}]%
        {Rahwan2019}
\bibfield{author}{\bibinfo{person}{Iyad Rahwan}, \bibinfo{person}{Manuel
  Cebrian}, \bibinfo{person}{Nick Obradovich}, \bibinfo{person}{Josh Bongard},
  \bibinfo{person}{Jean~François Bonnefon}, \bibinfo{person}{Cynthia
  Breazeal}, \bibinfo{person}{Jacob~W. Crandall}, \bibinfo{person}{Nicholas~A.
  Christakis}, \bibinfo{person}{Iain~D. Couzin}, \bibinfo{person}{Matthew~O.
  Jackson}, \bibinfo{person}{Nicholas~R. Jennings}, \bibinfo{person}{Ece
  Kamar}, \bibinfo{person}{Isabel~M. Kloumann}, \bibinfo{person}{Hugo
  Larochelle}, \bibinfo{person}{David Lazer}, \bibinfo{person}{Richard
  McElreath}, \bibinfo{person}{Alan Mislove}, \bibinfo{person}{David~C.
  Parkes}, \bibinfo{person}{Alex~‘Sandy’ Pentland},
  \bibinfo{person}{Margaret~E. Roberts}, \bibinfo{person}{Azim Shariff},
  \bibinfo{person}{Joshua~B. Tenenbaum}, {and} \bibinfo{person}{Michael
  Wellman}.} \bibinfo{year}{2019}\natexlab{}.
\newblock \showarticletitle{{Machine behaviour}}.
\newblock \bibinfo{journal}{\emph{Nature}} \bibinfo{volume}{568},
  \bibinfo{number}{7753} (\bibinfo{year}{2019}), \bibinfo{pages}{477--486}.
\newblock
\showISSN{14764687}
\urldef\tempurl%
\url{https://doi.org/10.1038/s41586-019-1138-y}
\showDOI{\tempurl}


\bibitem[\protect\citeauthoryear{Ramakrishnan, Kamar, Nushi, Dey, Shah, and
  Horvitz}{Ramakrishnan et~al\mbox{.}}{2019}]%
        {Ramakrishnan2019}
\bibfield{author}{\bibinfo{person}{Ramya Ramakrishnan}, \bibinfo{person}{Ece
  Kamar}, \bibinfo{person}{Besmira Nushi}, \bibinfo{person}{Debadeepta Dey},
  \bibinfo{person}{Julie Shah}, {and} \bibinfo{person}{Eric Horvitz}.}
  \bibinfo{year}{2019}\natexlab{}.
\newblock \showarticletitle{{Overcoming Blind Spots in the Real World:
  Leveraging Complementary Abilities for Joint Execution}}.
\newblock \bibinfo{journal}{\emph{Proceedings of the AAAI Conference on
  Artificial Intelligence}}  \bibinfo{volume}{33} (\bibinfo{year}{2019}),
  \bibinfo{pages}{6137--6145}.
\newblock
\showISSN{2159-5399}
\urldef\tempurl%
\url{https://doi.org/10.1609/aaai.v33i01.33016137}
\showDOI{\tempurl}


\bibitem[\protect\citeauthoryear{Rastkar, Murphy, and Murray}{Rastkar
  et~al\mbox{.}}{2014}]%
        {Rastkar2014AutomaticReports}
\bibfield{author}{\bibinfo{person}{Sarah Rastkar}, \bibinfo{person}{Gail~C
  Murphy}, {and} \bibinfo{person}{Gabriel Murray}.}
  \bibinfo{year}{2014}\natexlab{}.
\newblock \showarticletitle{{Automatic summarization of bug reports}}.
\newblock \bibinfo{journal}{\emph{IEEE Transactions on Software Engineering}}
  \bibinfo{volume}{40}, \bibinfo{number}{4} (\bibinfo{year}{2014}),
  \bibinfo{pages}{366--380}.
\newblock
\showISSN{00985589}
\urldef\tempurl%
\url{https://doi.org/10.1109/TSE.2013.2297712}
\showDOI{\tempurl}


\bibitem[\protect\citeauthoryear{Ratner, Sa, Wu, Selsam, and R\'{e}}{Ratner
  et~al\mbox{.}}{2016}]%
        {Ratner2016}
\bibfield{author}{\bibinfo{person}{Alexander Ratner},
  \bibinfo{person}{Christopher~De Sa}, \bibinfo{person}{Sen Wu},
  \bibinfo{person}{Daniel Selsam}, {and} \bibinfo{person}{Christopher R\'{e}}.}
  \bibinfo{year}{2016}\natexlab{}.
\newblock \showarticletitle{Data Programming: Creating Large Training Sets,
  Quickly}. In \bibinfo{booktitle}{\emph{Proceedings of the 30th International
  Conference on Neural Information Processing Systems}} (Barcelona, Spain)
  \emph{(\bibinfo{series}{NIPS'16})}. \bibinfo{publisher}{Curran Associates
  Inc.}, \bibinfo{address}{Red Hook, NY, USA}, \bibinfo{pages}{3574–3582}.
\newblock
\showISBNx{9781510838819}


\bibitem[\protect\citeauthoryear{Recht, Roelofs, Schmidt, and Shankar}{Recht
  et~al\mbox{.}}{2019}]%
        {Recht2019}
\bibfield{author}{\bibinfo{person}{Benjamin Recht}, \bibinfo{person}{Rebecca
  Roelofs}, \bibinfo{person}{Ludwig Schmidt}, {and} \bibinfo{person}{Vaishaal
  Shankar}.} \bibinfo{year}{2019}\natexlab{}.
\newblock \showarticletitle{{Do ImageNet classifiers generalize to ImageNet?}}
\newblock \bibinfo{journal}{\emph{36th International Conference on Machine
  Learning, ICML 2019}}  \bibinfo{volume}{2019-June} (\bibinfo{year}{2019}),
  \bibinfo{pages}{9413--9424}.
\newblock
\showISBNx{9781510886988}


\bibitem[\protect\citeauthoryear{Ren, Amershi, Lee, Suh, and Williams}{Ren
  et~al\mbox{.}}{2017}]%
        {Ren2017}
\bibfield{author}{\bibinfo{person}{Donghao Ren}, \bibinfo{person}{Saleema
  Amershi}, \bibinfo{person}{Bongshin Lee}, \bibinfo{person}{Jina Suh}, {and}
  \bibinfo{person}{Jason~D. Williams}.} \bibinfo{year}{2017}\natexlab{}.
\newblock \showarticletitle{{Squares: Supporting Interactive Performance
  Analysis for Multiclass Classifiers}}.
\newblock \bibinfo{journal}{\emph{IEEE Transactions on Visualization and
  Computer Graphics}} \bibinfo{volume}{23}, \bibinfo{number}{1}
  (\bibinfo{year}{2017}), \bibinfo{pages}{61--70}.
\newblock
\showISSN{10772626}
\urldef\tempurl%
\url{https://doi.org/10.1109/TVCG.2016.2598828}
\showDOI{\tempurl}


\bibitem[\protect\citeauthoryear{Rose, Engel, Cramer, and Cowley}{Rose
  et~al\mbox{.}}{2010}]%
        {Rose2010}
\bibfield{author}{\bibinfo{person}{Stuart Rose}, \bibinfo{person}{Dave Engel},
  \bibinfo{person}{Nick Cramer}, {and} \bibinfo{person}{Wendy Cowley}.}
  \bibinfo{year}{2010}\natexlab{}.
\newblock \showarticletitle{{Automatic Keyword Extraction from Individual
  Documents}}.
\newblock \bibinfo{journal}{\emph{Text Mining: Applications and Theory}}
  \bibinfo{number}{March} (\bibinfo{year}{2010}), \bibinfo{pages}{1--20}.
\newblock
\showISBNx{9780470749821}
\urldef\tempurl%
\url{https://doi.org/10.1002/9780470689646.ch1}
\showDOI{\tempurl}


\bibitem[\protect\citeauthoryear{Schr{\"{o}}ter, Bettenburg, and
  Premraj}{Schr{\"{o}}ter et~al\mbox{.}}{2010}]%
        {Schroter2010DoBugs}
\bibfield{author}{\bibinfo{person}{Adrian Schr{\"{o}}ter},
  \bibinfo{person}{Nicolas Bettenburg}, {and} \bibinfo{person}{Rahul Premraj}.}
  \bibinfo{year}{2010}\natexlab{}.
\newblock \showarticletitle{{Do stack traces help developers fix bugs?}}. In
  \bibinfo{booktitle}{\emph{Proceedings - International Conference on Software
  Engineering}}. \bibinfo{pages}{118--121}.
\newblock
\showISBNx{9781424468034}
\showISSN{02705257}
\urldef\tempurl%
\url{https://doi.org/10.1109/MSR.2010.5463280}
\showDOI{\tempurl}


\bibitem[\protect\citeauthoryear{Selbst, Boyd, Friedler, Venkatasubramanian,
  and Vertesi}{Selbst et~al\mbox{.}}{2019}]%
        {Selbst2019}
\bibfield{author}{\bibinfo{person}{Andrew~D. Selbst}, \bibinfo{person}{Danah
  Boyd}, \bibinfo{person}{Sorelle~A. Friedler}, \bibinfo{person}{Suresh
  Venkatasubramanian}, {and} \bibinfo{person}{Janet Vertesi}.}
  \bibinfo{year}{2019}\natexlab{}.
\newblock \showarticletitle{{Fairness and abstraction in sociotechnical
  systems}}.
\newblock \bibinfo{journal}{\emph{FAT* 2019 - Proceedings of the 2019
  Conference on Fairness, Accountability, and Transparency}}
  (\bibinfo{year}{2019}), \bibinfo{pages}{59--68}.
\newblock
\showISBNx{9781450361255}
\urldef\tempurl%
\url{https://doi.org/10.1145/3287560.3287598}
\showDOI{\tempurl}


\bibitem[\protect\citeauthoryear{Shrinivasan and Van~Wijk}{Shrinivasan and
  Van~Wijk}{2008}]%
        {Shrinivasan2008}
\bibfield{author}{\bibinfo{person}{Yedendra~B. Shrinivasan} {and}
  \bibinfo{person}{Jarke~J. Van~Wijk}.} \bibinfo{year}{2008}\natexlab{}.
\newblock \showarticletitle{{Supporting the analytical reasoning process in
  information visualization}}.
\newblock \bibinfo{journal}{\emph{Conference on Human Factors in Computing
  Systems - Proceedings}} (\bibinfo{year}{2008}), \bibinfo{pages}{1237--1246}.
\newblock
\showISBNx{9781605580111}
\urldef\tempurl%
\url{https://doi.org/10.1145/1357054.1357247}
\showDOI{\tempurl}


\bibitem[\protect\citeauthoryear{Simonite}{Simonite}{2018}]%
        {simonite2018comes}
\bibfield{author}{\bibinfo{person}{Tom Simonite}.}
  \bibinfo{year}{2018}\natexlab{}.
\newblock \showarticletitle{When it comes to gorillas, google photos remains
  blind}.
\newblock \bibinfo{journal}{\emph{Wired, January}}  \bibinfo{volume}{13}
  (\bibinfo{year}{2018}).
\newblock


\bibitem[\protect\citeauthoryear{Sun, Lo, Khoo, and Jiang}{Sun
  et~al\mbox{.}}{2011}]%
        {Sun2011}
\bibfield{author}{\bibinfo{person}{Chengnian Sun}, \bibinfo{person}{David Lo},
  \bibinfo{person}{Siau~Cheng Khoo}, {and} \bibinfo{person}{Jing Jiang}.}
  \bibinfo{year}{2011}\natexlab{}.
\newblock \showarticletitle{{Towards more accurate retrieval of duplicate bug
  reports}}.
\newblock \bibinfo{journal}{\emph{2011 26th IEEE/ACM International Conference
  on Automated Software Engineering, ASE 2011, Proceedings}}
  (\bibinfo{year}{2011}), \bibinfo{pages}{253--262}.
\newblock
\showISBNx{9781457716393}
\urldef\tempurl%
\url{https://doi.org/10.1109/ASE.2011.6100061}
\showDOI{\tempurl}


\bibitem[\protect\citeauthoryear{Sun, Lo, Wang, Jiang, and Khoo}{Sun
  et~al\mbox{.}}{2010}]%
        {Sun2010}
\bibfield{author}{\bibinfo{person}{Chengnian Sun}, \bibinfo{person}{David Lo},
  \bibinfo{person}{Xiaoyin Wang}, \bibinfo{person}{Jing Jiang}, {and}
  \bibinfo{person}{Siau~Cheng Khoo}.} \bibinfo{year}{2010}\natexlab{}.
\newblock \showarticletitle{{A discriminative model approach for accurate
  duplicate bug report retrieval}}.
\newblock \bibinfo{journal}{\emph{Proceedings - International Conference on
  Software Engineering}}  \bibinfo{volume}{1} (\bibinfo{year}{2010}),
  \bibinfo{pages}{45--54}.
\newblock
\showISBNx{9781605587196}
\showISSN{02705257}
\urldef\tempurl%
\url{https://doi.org/10.1145/1806799.1806811}
\showDOI{\tempurl}


\bibitem[\protect\citeauthoryear{Wachter, Mittelstadt, and Russell}{Wachter
  et~al\mbox{.}}{2017}]%
        {Wachter2017}
\bibfield{author}{\bibinfo{person}{Sandra Wachter}, \bibinfo{person}{Brent
  Mittelstadt}, {and} \bibinfo{person}{Chris Russell}.}
  \bibinfo{year}{2017}\natexlab{}.
\newblock \showarticletitle{{Counterfactual Explanations Without Opening the
  Black Box: Automated Decisions and the GDPR}}.
\newblock \bibinfo{journal}{\emph{SSRN Electronic Journal}}
  (\bibinfo{year}{2017}), \bibinfo{pages}{1--52}.
\newblock
\showISSN{1556-5068}
\urldef\tempurl%
\url{https://doi.org/10.2139/ssrn.3063289}
\showDOI{\tempurl}


\bibitem[\protect\citeauthoryear{Wakabayashi}{Wakabayashi}{2018}]%
        {wakabayashi2018self}
\bibfield{author}{\bibinfo{person}{Daisuke Wakabayashi}.}
  \bibinfo{year}{2018}\natexlab{}.
\newblock \showarticletitle{Self-driving Uber car kills pedestrian in Arizona,
  where robots roam}.
\newblock \bibinfo{journal}{\emph{The New York Times}}  \bibinfo{volume}{3}
  (\bibinfo{year}{2018}), \bibinfo{pages}{19}.
\newblock


\bibitem[\protect\citeauthoryear{Weick}{Weick}{1995}]%
        {Weick1995}
\bibfield{author}{\bibinfo{person}{Karl~E Weick}.}
  \bibinfo{year}{1995}\natexlab{}.
\newblock \showarticletitle{{Sensemaking in Organizations (Foundations for
  Organizational Science)}}.
\newblock \bibinfo{journal}{\emph{Star}} (\bibinfo{year}{1995}).
\newblock
\showISBNx{080397177X}
\showISSN{0956-5221}


\bibitem[\protect\citeauthoryear{Wu, Ribeiro, Heer, and Weld}{Wu
  et~al\mbox{.}}{2019}]%
        {Wu2019}
\bibfield{author}{\bibinfo{person}{Tongshuang Wu}, \bibinfo{person}{Marco~Tulio
  Ribeiro}, \bibinfo{person}{Jeffrey Heer}, {and} \bibinfo{person}{Daniel
  Weld}.} \bibinfo{year}{2019}\natexlab{}.
\newblock \showarticletitle{{{\{}E{\}}rrudite: Scalable, Reproducible, and
  Testable Error Analysis}}.
\newblock \bibinfo{journal}{\emph{Proceedings of the 57th Conference of the
  Association for Computational Linguistics}} (\bibinfo{year}{2019}),
  \bibinfo{pages}{747--763}.
\newblock
\urldef\tempurl%
\url{https://www.aclweb.org/anthology/P19-1073}
\showURL{%
\tempurl}


\bibitem[\protect\citeauthoryear{Yeasmin, Roy, and Schneider}{Yeasmin
  et~al\mbox{.}}{2014}]%
        {Yeasmin2014}
\bibfield{author}{\bibinfo{person}{Shamima Yeasmin},
  \bibinfo{person}{Chanchal~K. Roy}, {and} \bibinfo{person}{Kevin~A.
  Schneider}.} \bibinfo{year}{2014}\natexlab{}.
\newblock \showarticletitle{{Interactive visualization of bug reports using
  topic evolution and extractive summaries}}.
\newblock \bibinfo{journal}{\emph{Proceedings - 30th International Conference
  on Software Maintenance and Evolution, ICSME 2014}} (\bibinfo{year}{2014}),
  \bibinfo{pages}{421--425}.
\newblock
\showISBNx{9780769553030}
\urldef\tempurl%
\url{https://doi.org/10.1109/ICSME.2014.66}
\showDOI{\tempurl}


\bibitem[\protect\citeauthoryear{Zhang, Wang, Hao, Xie, Zhang, and Mei}{Zhang
  et~al\mbox{.}}{2015}]%
        {Zhang2015}
\bibfield{author}{\bibinfo{person}{Jie Zhang}, \bibinfo{person}{Xiao~Yin Wang},
  \bibinfo{person}{Dan Hao}, \bibinfo{person}{Bing Xie}, \bibinfo{person}{Lu
  Zhang}, {and} \bibinfo{person}{Hong Mei}.} \bibinfo{year}{2015}\natexlab{}.
\newblock \showarticletitle{{A survey on bug-report analysis}}.
\newblock \bibinfo{journal}{\emph{Science China Information Sciences}}
  \bibinfo{volume}{58}, \bibinfo{number}{2} (\bibinfo{year}{2015}),
  \bibinfo{pages}{1--24}.
\newblock
\showISSN{1674733X}
\urldef\tempurl%
\url{https://doi.org/10.1007/s11432-014-5241-2}
\showDOI{\tempurl}


\bibitem[\protect\citeauthoryear{Zimmermann, Premraj, Bettenburg, Just,
  Schr{\"{o}}ter, and Weiss}{Zimmermann et~al\mbox{.}}{2010}]%
        {Zimmermann2010}
\bibfield{author}{\bibinfo{person}{Thomas Zimmermann}, \bibinfo{person}{Rahul
  Premraj}, \bibinfo{person}{Nicolas Bettenburg}, \bibinfo{person}{Sascha
  Just}, \bibinfo{person}{Adrian Schr{\"{o}}ter}, {and}
  \bibinfo{person}{Cathrin Weiss}.} \bibinfo{year}{2010}\natexlab{}.
\newblock \showarticletitle{{What makes a good bug report?}}
\newblock \bibinfo{journal}{\emph{IEEE Transactions on Software Engineering}}
  \bibinfo{volume}{36}, \bibinfo{number}{5} (\bibinfo{year}{2010}),
  \bibinfo{pages}{618--643}.
\newblock
\showISBNx{9781595939951}
\showISSN{00985589}
\urldef\tempurl%
\url{https://doi.org/10.1109/TSE.2010.63}
\showDOI{\tempurl}


\end{thebibliography}
\end{document}